\pdfoutput=1
\documentclass[apj,numberedappendix,iop]{emulateapj}

\usepackage[utf8]{inputenc}
\usepackage[T1]{fontenc} 
\usepackage{microtype}
\usepackage{amsmath}
\usepackage{xcolor}
\usepackage{url}
\usepackage{hyperref}
\hypersetup{colorlinks=True, linkcolor=blue!50!black, citecolor=black,
  urlcolor=blue!50!black}

\usepackage[varg]{txfonts}
\providecommand\uptheta{\theta}
\providecommand\upmu{\mu}
\usepackage{graphicx}
\graphicspath{
  {figs/},
}
\setkeys{Gin}{width=\linewidth, height=0.85\textheight, keepaspectratio=true} 

\newcommand\Ion[2]{\ensuremath{\mathrm{#1\,\scriptstyle #2}}}
\newcounter{ionstage}
\renewcommand{\ion}[2]{
  \setcounter{ionstage}{#2}%
  \Ion{#1}{\Roman{ionstage}}}
\newcommand\ION[2]{\ensuremath{\mathrm{#1^{#2}}}}

\newcommand\NI{\ion{N}{1}}
\newcommand\OI{\ion{O}{1}}
\newcommand\Hbeta{\ensuremath{\mathrm{H}\beta}}

\newcommand\htwo{\ensuremath{\mathrm{H_2}}}
\newcommand\h[1]{\ensuremath{\mathrm{H^{#1}}}}
\newcommand\hplus{\ION{H}{+}}
\newcommand\hzero{\ION{H}{0}}
\newcommand\nzero{\ION{N}{0}}

\newcommand\U[1]{\ensuremath{\mathrm{#1}}}

\newcommand\kmps{\U{km\ s^{-1}}}
\newcommand\pcc{\U{cm^{-3}}}
\newcommand\K{\U{K}}

\newcommand\Ufluxphot{\U{photons\ s^{-1}\ cm^{-2}}}
\newcommand\Uintphot{\U{photons\ s^{-1}\ cm^{-2}\ sr^{-1}}}
\renewcommand\micron{\U{\upmu m}}

\renewcommand\ni{\Ion{N}{I}} 
\newcommand\nii{\Ion{N}{II}}
\newcommand\oi{\Ion{O}{I}}
\newcommand\oii{\Ion{O}{II}}
\newcommand\oiii{\Ion{O}{III}}

\newcommand\sii{\Ion{S}{II}}

\newcommand\cliii{\Ion{Cl}{III}}

\newcommand{\Cloudy}{\textsc{Cloudy}}
\newcommand\hii{\Ion{H}{II}}

\newcommand\hb{\Hbeta}

\newcommand\EW{\ensuremath{\mathrm{EW}}}
\newcommand\EWhb{\EW(\Hbeta)}
\newcommand\EWni{\ensuremath{\EW([\ni], \mathrm{Corr})}}
\newcommand\EWhbo{\ensuremath{\EW(\Hbeta, \mathrm{Obs})}}
\newcommand\EWhbc{\ensuremath{\EW(\Hbeta, \mathrm{Corr})}}
\newcommand\EWhbi{\ensuremath{\EW(\Hbeta, \mathrm{Int})}}
\newcommand\EWint{\ensuremath{\EW(\mathrm{Int})}}

\newcommand\thC{\(\uptheta^1\)~Ori~C}
\newcommand\thori[2]{\(\uptheta^{#1}\)~Ori~#2\@}
\newcommand\albedo{\ensuremath{\varpi}}
\newcommand\dust{\ensuremath{_{\mathrm{\scriptscriptstyle dust}}}} 

\newcommand\B{\ensuremath{_{\mathrm{B}}}}
\newcommand\proton{\ensuremath{_{\mathrm{p}}}}
\newcommand\electron{\ensuremath{_{\mathrm{e}}}}
\newcommand\Nline{\ensuremath{N_{\mathrm{line}}}}
\newcommand\euv{\ensuremath{_{\scriptscriptstyle\mathrm{EUV}}}}
\newcommand\fuv{\ensuremath{_{\scriptscriptstyle\mathrm{FUV}}}}
\newcommand\opni{\ensuremath{_{\scriptscriptstyle 5199}}}
\newcommand\SED{\ensuremath{\langle\lambda L_\lambda\rangle}}
\newcommand\band{\ensuremath{_{\mathrm{band}}}}  
\newcommand\eff{\ensuremath{_{\mathrm{eff}}}}  
\newcommand\mean[1]{\left\langle#1\right\rangle}
\newcommand\pump{\ensuremath{_{\mathrm{pump}}}}
\newcommand\scat{\ensuremath{_\mathrm{s}}}
\newcommand\vis{\ensuremath{_\mathrm{\scriptscriptstyle vis}}}

\makeatletter
\def\fps@figure{tbhp}
\def\fps@table{tbhp}
\makeatother

\begin{document} 
\title{Pumping up the [\NI{}] nebular lines}
\shorttitle{Pumping up the [\NI{}] nebular lines}
\author{
G. J. Ferland\altaffilmark{1},
W. J. Henney\altaffilmark{2}
C. R. O'Dell\altaffilmark{3},
R.L. Porter\altaffilmark{1},
P. A. M. van Hoof\altaffilmark{4}
and R.J.R. Williams\altaffilmark{5}}
\shortauthors{Ferland et al.}

\altaffiltext{1}{Department of Physics, Universi
ty of Kentucky, Lexington,
KY 40506, USA.}

\altaffiltext{2}{Centro de Radioastronomía y Astrofísica, UNAM Campus Morelia, Apartado Postal 3-72, 58090 Morelia, Michoacán, Mexico}

\altaffiltext{3}{Department of Physics and Astronomy, Vanderbilt University, Box 1807-B, Nashville, TN 37235.}

\altaffiltext{4}{Royal Observatory of Belgium, Ringlaan 3, 1180 Brussels,
Belgium}

\altaffiltext{5}{AWE plc, Aldermaston, Reading RG7 4PR}

\begin{abstract}
The optical [N I] doublet near 5200\AA\ is anomalously strong 
in a variety of emission-line objects. 
We compute a detailed photoionization model and use it to show 
that pumping by far-ultraviolet (FUV) stellar radiation previously 
posited as a general explanation applies to the Orion Nebula (M42) 
and its companion M43; but, 
it is unlikely to explain planetary nebulae and supernova remnants. 
Our models establish that the observed nearly constant equivalent width of 
[N I] with respect to the dust-scattered stellar continuum 
depends primarily on three factors: 
the FUV to visual-band flux ratio of the stellar population; 
the optical properties of the dust; 
and the line broadening where the pumping occurs. 
In contrast, the intensity ratio [N I]/\hb\ depends primarily on the 
FUV to extreme-ultraviolet ratio, 
which varies strongly with the spectral type of the exciting star. 
This is consistent with the observed difference of a factor of five 
between M42 and M43, which are excited by an O7 and B0.5 star respectively.
We derive a non-thermal broadening of order $5\ \kmps$ for the [N I] 
pumping zone and show that the broadening mechanism must be 
different from the large-scale turbulent motions that have been suggested 
to explain the line-widths in this H II region. 
A mechanism is required that operates at scales of a few astronomical units, 
which may be driven by thermal instabilities of neutral gas in the range 
1000 to 3000~K.
In an appendix, we describe how collisional and radiative processes 
are treated in the detailed model N I atom now included in the Cloudy plasma code.

\end{abstract}

\section{Introduction}
\label{sec:introduction}

The optical emission-line spectrum of a photoionized cloud has prominent recombination lines (\Ion{H}{I}, \Ion{He}{I}, and \Ion{He}{II}) and collisionally excited lines (forbidden lines such as [\oiii], [\oii], [\nii] and [\sii]).  
The forbidden lines are produced by ions that exist within the \hplus{} region, where the gas kinetic temperature is high enough (\(\sim 10^{4}\)~K) for the lines to be collisionally excited (\citealp{Osterbrock.D06Astrophysics-of-gaseous-nebulae-and-active}, hereafter AGN3).
Ions with potentials smaller than \hzero{} exist mainly in the 
photodissociation region (PDR), a cold (\(T \leq 10^3~\K\)) region beyond the \hplus--\,\hzero{} ionization front which are shielded from ionizing radiation.
The PDR does not produce strong optical emission due to its low temperature.

The [\NI{}] doublet at $\lambda \lambda 5199~\AA$ is an interesting exception to this rule.  
Atomic nitrogen has an ionization potential only slightly larger than that of hydrogen, 14.5~eV for \nzero, as opposed to 13.6~eV for \hzero\ \citep{Moore.C93-Tables_spectra_H_C_N_O}.  
These, together with the relatively slow charge exchange reactions between H and N \citep{Kingdon.J96Rate-Coefficients-for-Charge-Transfer}, mean that little \nzero{} is present in warm gas, 
so [\ni] has a small collisional contribution and the lines are generally weak.  
This expectation appears to be confirmed in high-resolution observations of nearby \hii{} regions such as Orion \citep{Baldwin.J00High-Resolution-Spectroscopy-of-Faint-Emission}, 
hereafter B2000, where the doublet has an observed intensity of only $3 \times 10^{-3}$ that of \Hbeta.
But we show in this paper that the ratio becomes higher within the 
central parts of the Orion Nebula, and much larger in the nearby
M43 nebula.

This study is motivated by the exceptionally strong intensity of the [\ni] doublet in several unusual classes of nebulae.  
Filaments in cool-core clusters of galaxies and filaments in the Crab Nebula can have the [\ni] doublet nearly as strong as \Hbeta{} \citep{Ferland.G09Collisional-heating-as-the-origin-of-filament, Davidson.K85Recent-developments-concerning-the-Crab}.  
The great [\ni] strength is the single most exceptional spectroscopic feature in the optical region for these nebulae, 
and could indicate that atomic gas has been heated to temperatures warm enough to collisionally excite the line.  
This could be done by a large flux of very hard photons or energetic particles, but is an area of active investigation.
Large-scale velocity variations within these objects could also enhance the
absorption of continuum photons, making continuum fluorescence more important.
The fact that several very different physical processes may be active
makes it difficult to understand what the strong [\NI{}] doublet tells us
about these unusual environments. 
It is, therefore, important to quantitatively explain these lines in the
arguably simplest case, an H~II region.

Continuum fluorescent excitation has been proposed to be 
an important contributor to the
intensity of the [\ni] doublet \citep{Bautista:1999}.
This process is unusual because the ground term of \nzero\ is not connected
to the upper levels of the observed [\ni] doublet by any LS-allowed transitions.
It is the breakdown of LS coupling in \ni\ which makes the process fast.
The FUV lines which pump the upper levels of the [\NI{}] doublet  
lie in the wavelength range
$\lambda \lambda 951\AA - 1161 \AA$.
The resulting intensity of optical [\ni] lines will depend on 
the atomic transition probabilities (a difficult atomic physics problem due
to the breakdown of LS coupling), the Spectral Energy Distribution (SED) of the incident stellar radiation field
around the $\lambda \lambda 951\AA - 1161 \AA$ driving lines, and 
gas motions in the region where continuum fluorescence occurs
since the driving lines become self shielded.
Appendix~\ref{sec:fluorescence} describes the fluorescence
mechanism in detail.

The purpose of this paper is to use the Orion star-forming region to check whether 
photoionization simulations can self consistently account for the observed [\ni] intensity.  
Orion is a relatively quiescent environment that can serve as a test bed for conventional nebular theory.  
Our simulations largely confirm the prediction by \citet{Bautista:1999} that the [\ni] lines are predominantly formed by continuum fluorescent excitation.  
We show that their intensity relative to \Hbeta{} is mainly set by the non-thermal component of line broadening in shallow regions of the PDR\@.  
The line broadening needed to account for the observed line intensities is consistent with that seen in Orion.  

The 5198, 5200\AA\ pair of lines are denoted as $\lambda 5199+$ in this paper.
These lines often appear as a single feature at low resolution or
when the intrinsic line widths are large.
Appendix~\ref{sec:NIdiagnostic} describes how the two lines
within $\lambda 5199+$ can be used to measure density if the 
[\ni] lines are collisionally excited.

\section{Observations}
\label{sec:observations}

In the \citet{Bautista:1999} study of [\ni] emission there was only a limited attempt to compare the results with observed line intensity ratios. 
This was done in a qualitative way for numerous planetary nebulae, supernova remnants, and Herbig-Haro objects, and it should be noted that the axes in his Figure~4 are all 100 times too large. 
The Orion Nebula (M~42, NGC~1976) presents an excellent opportunity for testing theories of the formation of [\ni] emission as the lines are known to be present under various conditions.

Fortunately, there is a recently published spectrophotometric study (\citealp{ODell:2010}, henceforth OH10) covering all of the brightest part of the Orion Nebula (the Huygens Region), the fainter outer region (the Extended Orion Nebula), and the nearby \hii{} region M~43 (NGC~1982).  
The inclusion of M~43 is particularly important since that object lies along the borderline between an object being a photoionized \hii{} region and its being a simple reflection nebula. 
This status is caused by the dominant star NU~Ori (spectral type B0.5, \citealp{ODell:2011}) being much cooler than the dominant ionizing star of M~42 (\thC{}, spectral type O7 V, \citealp{ODell:2011}). 
OH10 obtained moderate spectral resolution long slit samples at various distances from \thC{} and NU~Ori. 
Reddening corrections were determined for each spectrum. 
In additon to emission-line ratios relative to \Hbeta, absolute surface brightnesses in \Hbeta, were determined. 
An important measurement made in OH10 was that of the underlying continuum, the strength of this continuum being expressed as the equivalent width (\(\EWhb = I(\Hbeta)/I(\mathrm{Cont})\), where \(I(\Hbeta)\) is the surface brightness in the \Hbeta{} emission-line and \(I(\mathrm{Cont})\) is the surface brightness of the observed continuum per Angstrom). 
The units for \EWhb{} are \AA{}ngstroms (\AA).  
The expected \EWhb{} for the Huygens Region due to atomic processes is about 1700~\AA{} \citep{2001ARA&A..39...99O}.  
It has long been known \citep{Baldwin.J91Physical-conditions-in-the-Orion-Nebula} that the observed equivalent width (\EWhbo) is much smaller than this.
This indicates a strong scattered light component arises from Trapezium starlight backscattered by dust lying in the dense photon dominated region (PDR) that lies just beyond the ionized layer that separates \thC{} and the background Orion Molecular Cloud.  
OH10 demonstrate that \EWhbo{} decreases with increasing distance from \thC{} and that \EWhbo{} values for M~43 are comparable to the more distant samples within M~42. 
OH10 determined that the M~42 spectra beyond about 10\arcmin{} are increasingly affected by scattered light originating from the Huygens region. 
We have included only those samples from their ``inner'' region group with distances of less than 8\arcmin{} and all of their M~43 samples in this analysis.

\begin{figure*}
  \centering
  \includegraphics[trim=0 0 70 0, clip]{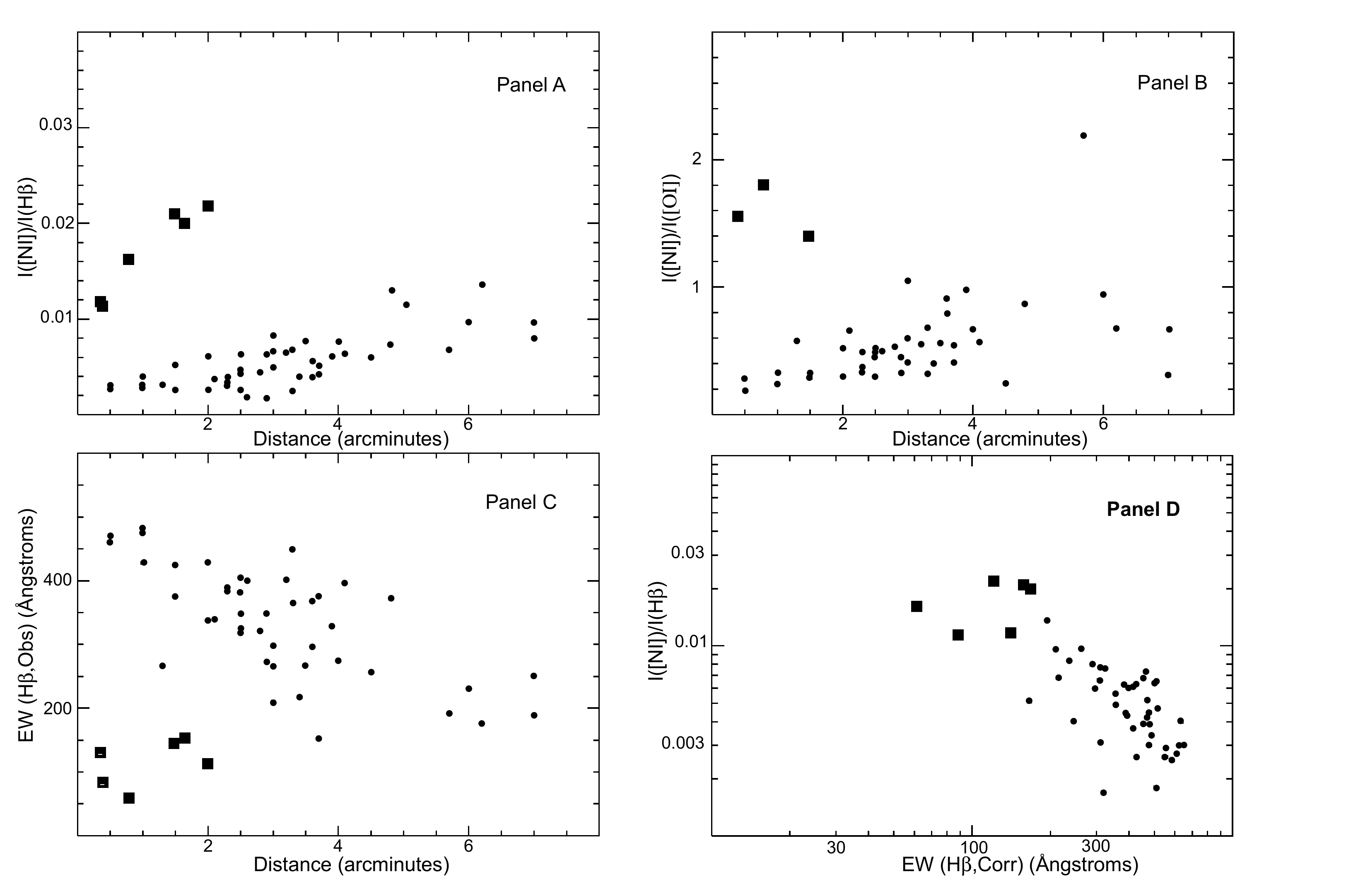}
  \caption{These four panels present the spectrophotometric results for the sample regions of M~42 and M~43 as described in the text.  Filled circles represent M~42 samples and filled squares represent M~43 samples.
The distances are from the center of the Trapezium for the filled circles representing M42 and from NU Ori for the filled squares representing M43. }
  \label{fig:spectrophotometry}
\end{figure*}

The reddening corrected results from OH10 are shown in Figure~\ref{fig:spectrophotometry}. 
Panel~A presents the reddening-corrected emission-line ratio \(I([\ni])/I(\Hbeta)\) as a function of distance from the dominant star (\thC{} for the M~42 results and NU~Ori for the M~43 results), where \(I([\ni])\) is the total emission from the forbidden \ION{N}{+} lines near 5200~\AA\@.  
Panel~B presents the ratio \(I([\ni])/I([\oi])\) (where \(I([\oi])\) is the sum of the neutral oxygen lines at 6300~\AA{} and 6363~\AA) as a function of distance. 
Panel~C presents \EWhbo{} as a function of distance.
Panel~D presents in logarithmic scale the \(I([\ni])/I(\Hbeta)\) ratio as a function of \EWhbc, where \EWhbc{} is the \EWhbo{} value corrected for the expected atomic continuum of 1700~\AA.

In Figure~\ref{fig:spectrophotometry}, panel~A we note that although there is a wide scatter, there is a general
increase in \(I([\ni])/I(\Hbeta)\) with increasing distance from \thC{}.
The M~43 line ratios are much larger and show an even more rapid increase with distance from NU~Ori.  
In panel B we note a small general increase in the \(I([\ni])/I([\oi])\) ratio with increasing distance from \thC{}, while the three samples for M~43 show ratios much larger than those for M~42.  
There are fewer samples for \(I([\oi])\) in M~43 because of its
much lower surface brightness. 
The most distant ratios in M~42 show a large scatter because of the difficulty in separating faint nebular emission from the strong foreground
night-sky [\oi] emission.
Panel~C shows that \EWhbo{} decreases markedly in M~42 (the scattered light continuum becomes stronger with increasing distance from \thC{}). 
The scattered light continuum is always stronger in the M~43 samples, but there is no obvious correlation with distance from NU~Ori.  
We should note here that the blister model for M~42 is well
established, so that we can expect a monotonic change in conditions when looking at lines of sight of greater distance. 
However, the physical model for M~43 is not established.
Is M~43 the simple Strömgren sphere with overlying foreground material in the east as suggested by its circular appearance or is it too a blister model object?

In Figure~\ref{fig:spectrophotometry}, panel~D we see that the the M~42 and M~43 samples form a well-defined sequence when considering the \(I([\ni])/I(\Hbeta)\) versus \EWhbc. 
A linear relation would indicate that the ratio \(I([\ni])/I(\mathrm{Cont,Corr})\) (which we will call \EWni)
is constant. 
\(I(\mathrm{Cont,Corr})\) is the observed continuum corrected for the atomic component. 
Considering the two nebulae separately, we calculate \EWni{} to be \(1.98\pm 0.65\) for M~42 and \(2.15 \pm 0.99\) for M~43. 
The presence of an approximate linear correlation suggests
that [\ni] emission is driven by non-ionizing continuum radiation. 
The value of \EWni{} can become a quantitative test for any suggested driving mechanism for the [\ni] emission and is pursued in the remainder of this paper.

\section{Predicted emission from a ray through inner regions of the Orion Nebula}
\label{sec:emission-from-ray}
This highest S/N observations are for bright inner regions of the Orion Nebula.
To quantify the various physical contributors to the formation of [\ni] lines we
recomputed the \citep{Baldwin.J91Physical-conditions-in-the-Orion-Nebula}
model of a ray through the the \hii\ region. 
Appendix~\ref{sec:NIAtomicModelAppendix} describes recent improvements in the treatment of \ni\ emission 
in the spectral simulation code Cloudy which we use to compute the spectrum.
The model is a layer in hydrostatic equilibrium: the outward stellar radiation pressure, 
largely due to grains, is balanced by gas,
turbulent, and magnetic pressures within the nebula.
The parameters are those given in BFM with the following exceptions:

\begin{itemize}
\item We include the five high-mass stars of the Trapezium (see Table~\ref{tab:stars}), using atmospheres from \citet{2003ApJS..146..417L} and 
\citet{Lanz.T07A-Grid-of-NLTE-Line-blanketed-Model}.  
This produces more 1000~\AA\
photons relative to the Lyman continuum than would be obtained from \thC{} alone.
\item We continue the calculation into the PDR and \htwo\ region, including the full
\htwo\ model described by \cite{Shaw.G05Molecular-Hydrogen-in-Star-forming-Regions:} and the chemistry network described by 
\citet{Abel.N05The-H-II-Region/PDR-Connection:-Self-consistent-Calculations}.
The calculation stops at a thickness corresponding to $A_V = 10^3$.
\item{We work in terms of stellar luminosities and the physical size of the
blister.  As a result the model is not plane parallel, it has a ratio of outer
to inner radius of about two.  
We simulate observing this structure by using the option to integrate
intensities along a pencil beam through the geometry.}
\item The gas is assumed to be in hydrostatic equilibrium, as in BFM.  
We include magnetic, but not turbulent, pressure in the gas equation of state.  
\item A ``tangled'' magnetic field is assumed, as described in Appendix~C of \citet{2005ApJ...621..328H}, with an effective magnetic adiabatic index of \(\gamma_\mathrm{mag} = 1.0\).  
The magnetic field in the ionized gas is chosen so as to give a ratio of gas pressure to magnetic pressure (plasma \(\beta\)) of 10, which is a typical value found for \hii{} regions \citep{1981ApJ...247L..77H, Harvey-Smith:2011, Rodriguez:2011}.  
Together with the assumption of \(\gamma_\mathrm{mag} = 1.0\), this implies a constant Alfvén speed of \(v_\mathrm{A} \simeq 3.5~\kmps\), which is roughly consistent with both numerical simulations \citep{Arthur:2011} and observational limits \citep{Crutcher:2010}.  The magnetic pressure and gas pressure are therefore roughly equal in the PDR (\(T \sim 1000~\K\), \(\beta \sim 1\)), whereas magnetic pressure dominates in the colder molecular gas (\(T \sim 100~\K\), \(\beta \ll 1\)).  
\end{itemize}

\begin{figure}
  \centering
  \includegraphics{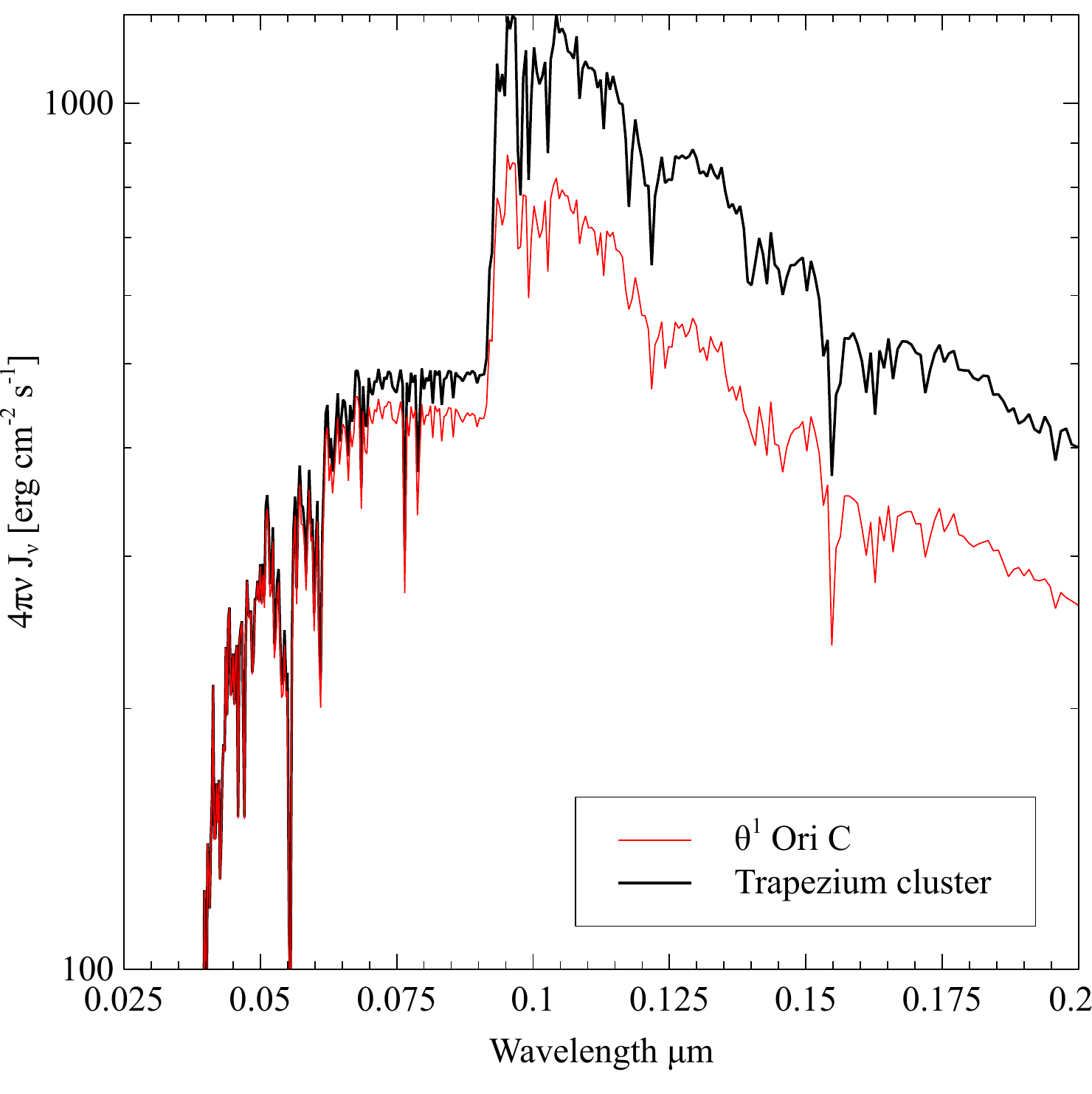}
  \caption{The SED of \thC\ is the lower curve while the
  heavier higher curve gives the SED of the Trapezium stars,
  using the stellar parameters summerized in Table \ref{tab:stars}
  and the predictions of \citet{2003ApJS..146..417L} and 
  \citet{Lanz.T07A-Grid-of-NLTE-Line-blanketed-Model}.
  The figure is centered on 0.1~\micron, which is 1000\AA.}
  \label{fig:SED}
\end{figure}

Table \ref{tab:stars} lists the stars we include.
Figure \ref{fig:SED} compares two SEDs.
The lower curve is \thC\ by itself
while the higher curve includes all stars.
The largest differences are in the intensity of the FUV relative to the
Lyman continuum.
In the case where [\ni] is photoexcited and \hb\ produced by recombination,
the line intensity ratio is proportional to the ratio of the FUV relative 
to the Lyman continuum.
The [\ni] pumping rate will depend on the intensity of the stellar radiation field
at the wavelengths of the FUV \ni\ lines.
Photospheric absorption lines are present across the FUV,
making an accurate stellar model essential.

\begin{table}
  \centering
  \caption{Massive stars in M42 and M43}
  \label{tab:stars}
  \begin{tabular}{lrlrrrl}
    \hline
    Star & \(M/M_\odot\) & SP type & \(\log L/L_\odot\) & \(T/\K\) & \(\log g\) & Refs \\
    \hline
    \multicolumn{7}{c}{\textit{M42 inner}} \\
    \thori{1}{A} & 14 & B0.5 V & 4.45 & 30,000 & 4.0 & 1 \\ 
    \thori{1}{B} & 7 & B3 V & 3.25 & 18,000 & 4.1 & 2 \\
    \thori{1}{C} & 32 & O7 V & 5.31 & 39,000 & 4.1 & 1, 3 \\
    \thori{1}{C2} & 12 & B1 IV & 4.20 & 25,000 & 3.9 & 3, 4 \\
    \thori{1}{D} & 18 & B0.5 V & 4.47 & 32,000 & 4.2 & 1 \\
    \hline
    \multicolumn{7}{c}{\textit{M42 outer}} \\
    \thori{2}{A} & 30 & O9 V & 4.93 & 35,000 & 4.0 & 1 \\ 
    \thori{2}{B} & 7 & B0.5 V & 4.11 & 29,000 & 4.1 & 1 \\
    \thori{2}{C} & 6 & B4 V & 3.00 & 17,000 & 4.1 & 5, 6 \\
    LP~Ori & 10 & B1.5 V & 3.75 & 23,000 & 4.1 & 5, 6 \\
    P1744  & 5  & B5 V & 2.70 & 16,000 & 4.1 & 5, 6 \\
    \hline
    \multicolumn{7}{c}{\textit{M43}} \\
    NU~Ori & 18 & B0.5 V & 4.42 & 31,000 & 4.2 & 7 \\
    \hline
  \end{tabular}
  \tablecomments{
    Stellar parameters of all stars more massive than \(5~M_\odot\) within the confines of M42 and M43, divided into three groups.  The ``\textit{M42 inner}'' group are the Trapezium stars, which excite the bright Huygens region of the Orion Nebula.  The ``\textit{M42 outer}'' group are situated 2--10\arcmin{} south of the Trapezium and contribute to the excitation of the Extended Orion Nebula.
  }
  \tablerefs{
    1.~\citet{2006A&A...448..351S};
    2.~\citet{Weigelt:1999};
    3.~\citet{Schertl:2003};
    4.~\citet{Lehmann:2010a};
    5.~\citet{Malkov:1992};
    6.~\citet{Fitzpatrick:2005};
    7.~\citet{Simon-Diaz:2011a}.
  }
\end{table}

The true atmosphere of \thC{} remains highly uncertain.  
The object is a close binary with an extended atmosphere and a detected and 
periodically variable magnetic field.  
In addition to periodic variations with a period of 15.4 days there are 
known non-periodic radial velocity and spectral variations.  
These characteristics are summarized in \citet{Stahl:2008}.  
The established complexity of the atmosphere means that predictions of the 
SED of simple atmosphere models have a corresponding uncertainty of undefined magnitude.

\pagebreak[4]
\subsection{Properties of the cloud}
\begin{figure}
  \centering
  \includegraphics{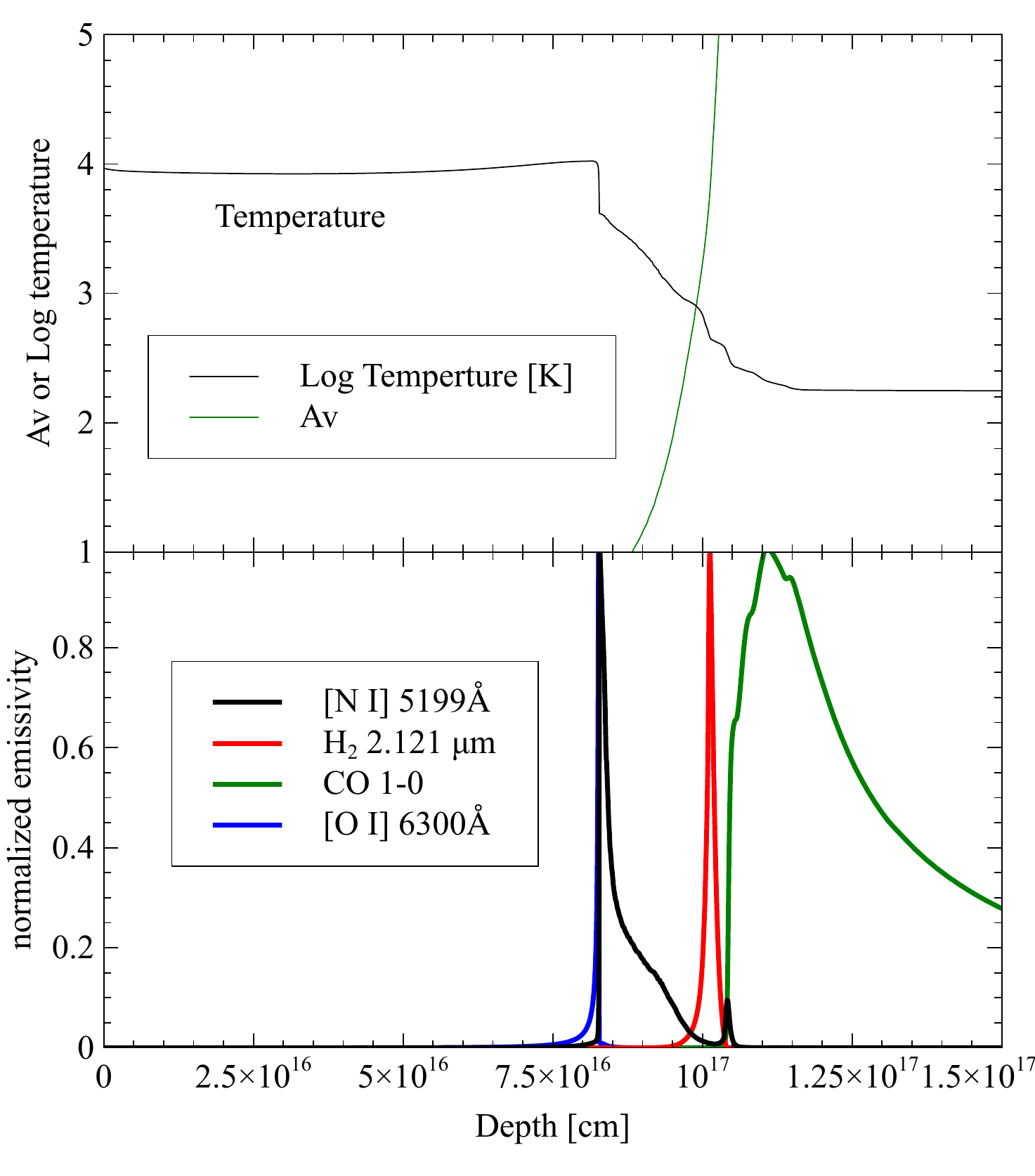}
  \caption{The temperature, extinction, 
  and emissivity of several important emission lines,
  are shown as a function of depth into the \hplus\ layer.
  The upper panel shows the log of the gas kinetic temperature
and the visual extinction A$_V$, with the values of both 
indicated on the left axis.  
The lower panel shows the normalized emissivity for several
  important lines. 
  This is the volume emissivity (erg cm$^{-3}$ s$^{-1}$)
  divided by the peak emissivity for each line to place them on the same scale.
 }
  \label{fig:EmissivityVsDepth}
\end{figure}

\nopagebreak[4]
The upper panel of Figure \ref{fig:EmissivityVsDepth} shows the temperature structure of the 
cloud along our ray.
The \hplus\ region has a temperature of around $10^4$ K while the gas kinetic temperature falls to around
300 K in the \h0\ region or PDR.
The deeper \htwo\ region is also colder.
The \hplus\ region is thicker than was found in 
\citet{Baldwin.J91Physical-conditions-in-the-Orion-Nebula} due to
magnetic support.

The lower panel of Figure \ref{fig:EmissivityVsDepth} shows the volume emissivity of the 
$\lambda 5199^+$ lines along this ray.  
For reference the lower panel also shows the emissivity of some well-observed
\htwo, CO, and [\oi] lines.
We see that both [\oi] and [\ni] lines form near the \hplus - \h0 ionization front, 
while the \htwo\ and CO lines form near the \h0 - \htwo\ dissociation front.

The \htwo\ line is mainly formed by continuum fluorescent excitation for
a PDR near an \hii\ region \citep{1985ApJ...291..722T}.
This formation processes is very similar to that forming the [\ni] lines.
The \htwo\ and [\ni] lines form at either edge of the PDR due to
a combination of abundance and FUV line optical depth effects.

\begin{figure}
  \centering
  \includegraphics{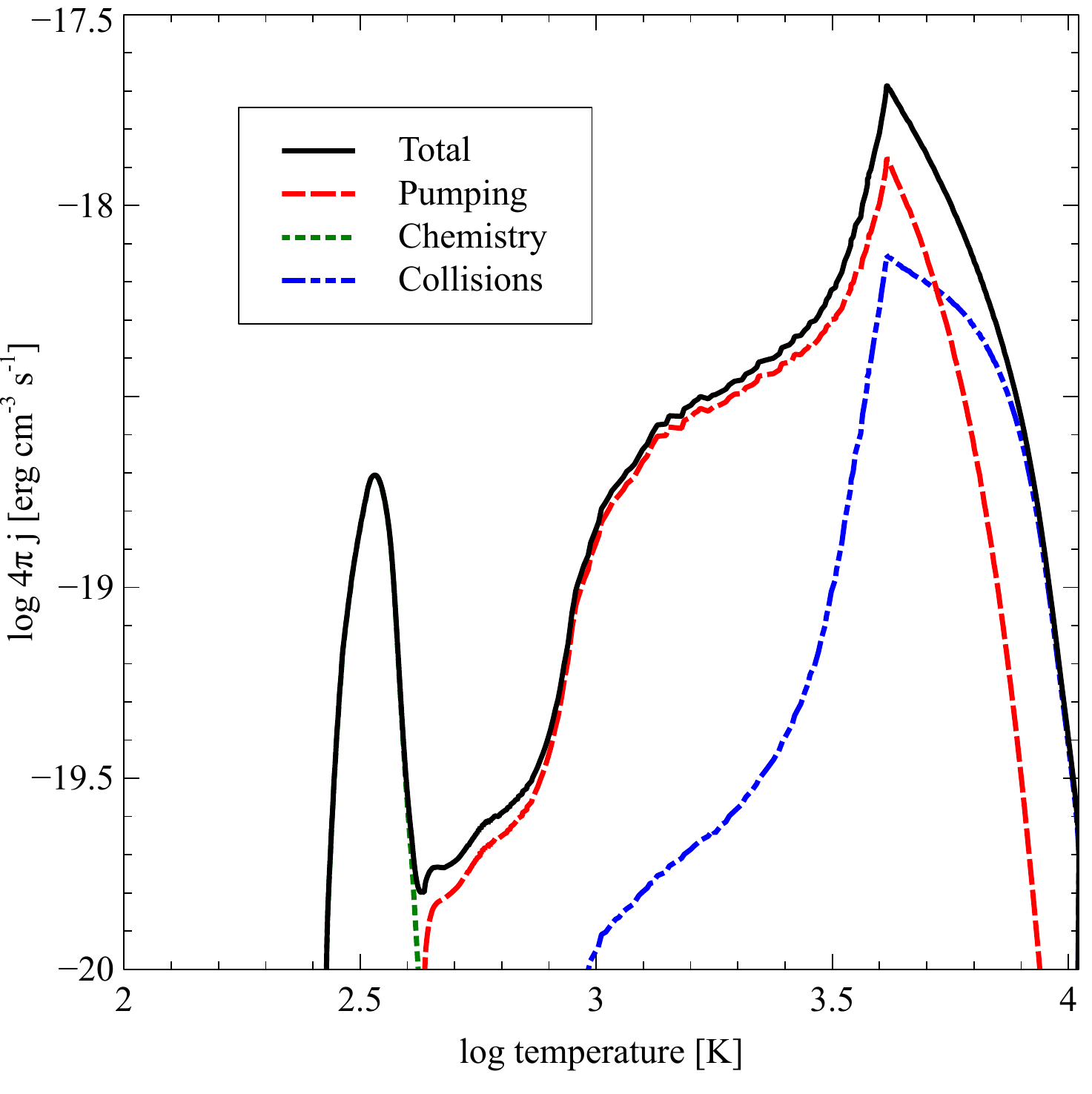}
  \caption{The volume emissivity of [\ni] $\lambda 5199^+$ vs gas kinetic temperature.
  This is a different view of the data in Figure \ref{fig:EmissivityVsDepth}.
  The total emissivity, and the contributions from collisions, continuum pumping, 
  and the
  chemical dissociation processes described in Appendix~\ref{sec:NIAtomicModelAppendix}, are shown.
  }
  \label{fig:EmissivityVsTemperature}
\end{figure}

Figure \ref{fig:EmissivityVsTemperature} shows the volume emissivity of the
[\ni] line as a function of gas kinetic temperature.
This is a convenient way to visualize the rapid changes in emissivity that
occur near the H$^+ - $H$^0$ ionization front, where both
emissivity and temperature change rapidly.
This does not indicate the total contribution of various processes to the observed
line since the surface brightness is the integral of the emissivity over the 
emitting volume.  
The size of each volume element changes dramatically as the conditions change.

\pagebreak[4]The peak emissivity occurs at a gas kinetic temperature of $\sim 4000$~K,
with contributions from continuum pumping and collisions.
The emissivity increases as the temperature decreases due to the increasing
N$^0$ abundance in cooler regions.
Continuum pumping increases with increasing N$^0$ abundance.
The emissivity decreases at lower temperatures due to increasing optical depths
in the FUV lines.
The rise in emissivity at temperatures $\sim 300 $K is due to formation
by molecular dissociation, using the estimates outlined in Appendix~\ref{sec:diss-excit}.
  
\subsection{The effects of non-thermal broadening}

\begin{figure}
  \centering
  \includegraphics{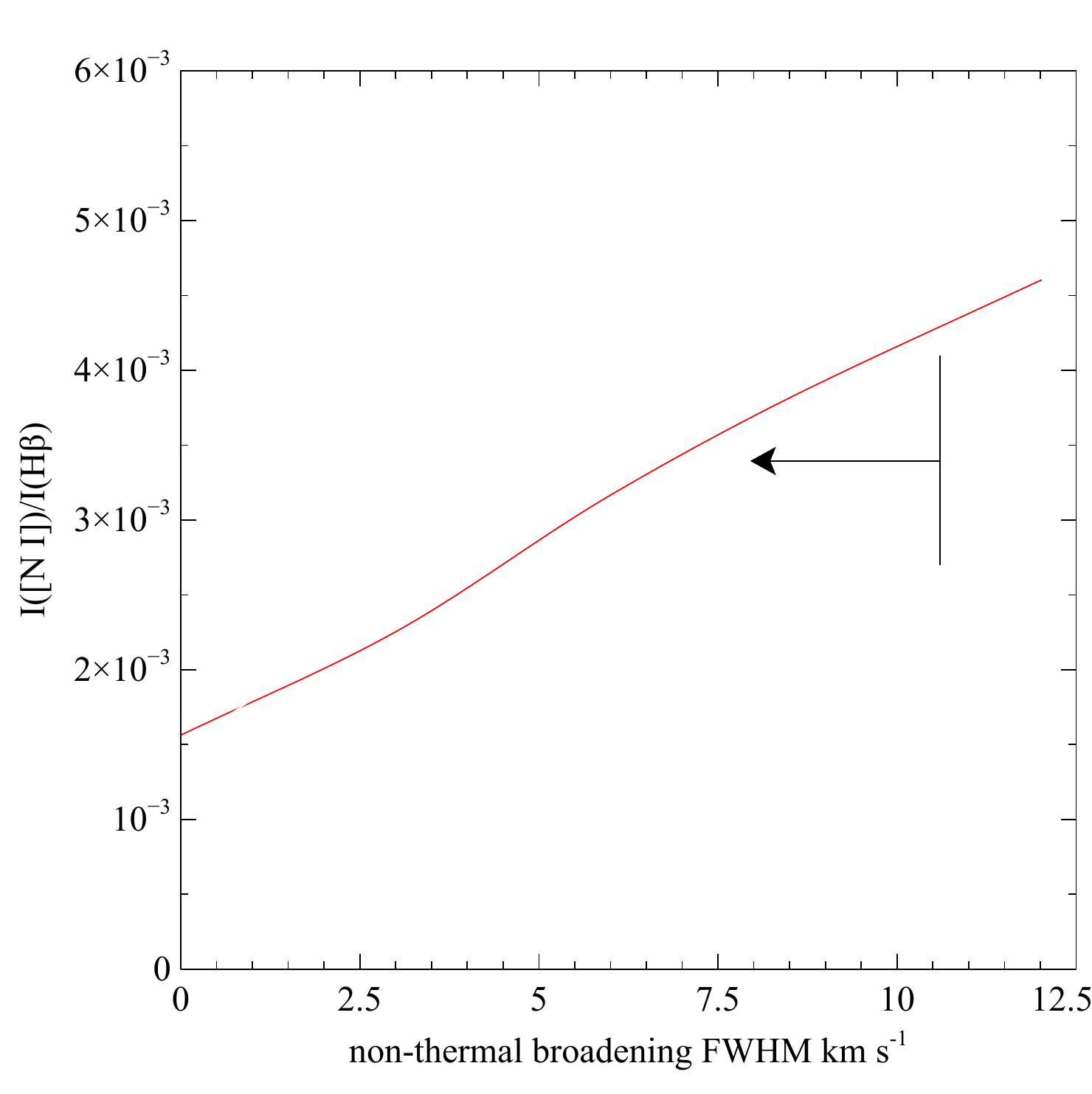}
  \caption{Surface brightness of the [\NI{}] line relative to \Hbeta{} as a function of the FWHM of the non-thermal line broadening component in the [\NI{}] formation region. 
  The observed ratio is indicated along with error bars that represent the scatter in the ratio.
  The observed FWHM is indicated and is really an upper limit to the
  microturbulent broadening, as discussed in Appendix~\ref{sec:interpr-equiv-widths}.
  }
  \label{fig:varyTurb}
\end{figure}

Given these assumptions the only free parameter is the turbulent contribution to the line width.
Figure \ref{fig:varyTurb} shows the predicted [\ni] 5199$^+$ / \hb\ 
intensity ratio as  function of the turbulence.
We have determined the broadening of the [\ni] lines from spectra made 
with the HIRES spectrograph as part of a program studying mass 
loss from proplyds in the Huygens Region 
\citep{Henney.W99A-Keck-High-Resolution-Spectroscopic-Study}. 
In the proplyd-free portions of these long-slit spectra, 
the average observed FWHM was \(13.65 \pm 1.91~\kmps\). 
The instrumental FWHM of the comparison lines was \(8.28 \pm 0.40~\kmps\). 
If the [\ni] emission arises from the region with \(T = 1000\) to \(3000~\K\), 
then the thermal component of the broadening would be \(2\) to \(3~\kmps\).
After quadratic subtraction of the instrumental and thermal widths 
from the observed FWHM, there is a residual non-thermal broadening component of 
\(10.6 \pm 1.9~\kmps\).  
The four proplyds in the sample (150-353, 170-337, 177-341, 182-413) 
have an average distance from \thC{} of \(0.56 \pm 0.29'\).
After examination of Panel A of Figure 1, we see that the expected ratio 
\(I([\ni])/I(\Hbeta)\) would be \(0.0034 \pm 0.0007\).  
These values are indicated
in Figure~\ref{fig:varyTurb} for comparison with the predictions of our model.  
The linewidth is given as an upper limit, 
since the observed emission profile will in general include contributions from macroscopic and microscopic broadening processes 
(see discussion in Appendix~\ref{sec:non-thermal-line}) 
whereas only the latter will contribute to the pumping efficiency.

In Appendix~\ref{sec:NIAtomicModelAppendix} we show that in the fluorescent scenario the strength of the [\NI{}] emission lines with respect to \Hbeta{} is close to linearly proportional to the degree of line broadening in the region in which the lines are pumped.  In order to reproduce the observed brightness, our model requires a FWHM for the broadening (assuming a Gaussian line profile) of \(\simeq 10~\kmps\).  
If this broadening were to be thermal, then a temperature of \(> 10,000~\K\) in the pumping region would be required, which is much larger than the \(\approx 2000~\K \) predicted by our Cloudy model.  
Instead, it is likely that the majority of the broadening 
is non-thermal in nature. 
Significant non-thermal line widths have been reported 
in the spectra of Orion Nebula 
emission-lines 
\citep{2001ARA&A..39...99O,2003AJ....125.2590O,2008RMxAA..44..181G}. 
The nature of
the processes producing this broadening is not known, 
but it must be important as its 
magnitude indicates that as much energy is contained 
there as is contained in the 
components explained by basic photo-ionization physics. 


\section{Discussion}
\label{sec:discussion}

\subsection{Spatial variation of intensity ratios}
\label{sec:spat-vari-intens}
\begin{figure}
  \centering
  \includegraphics{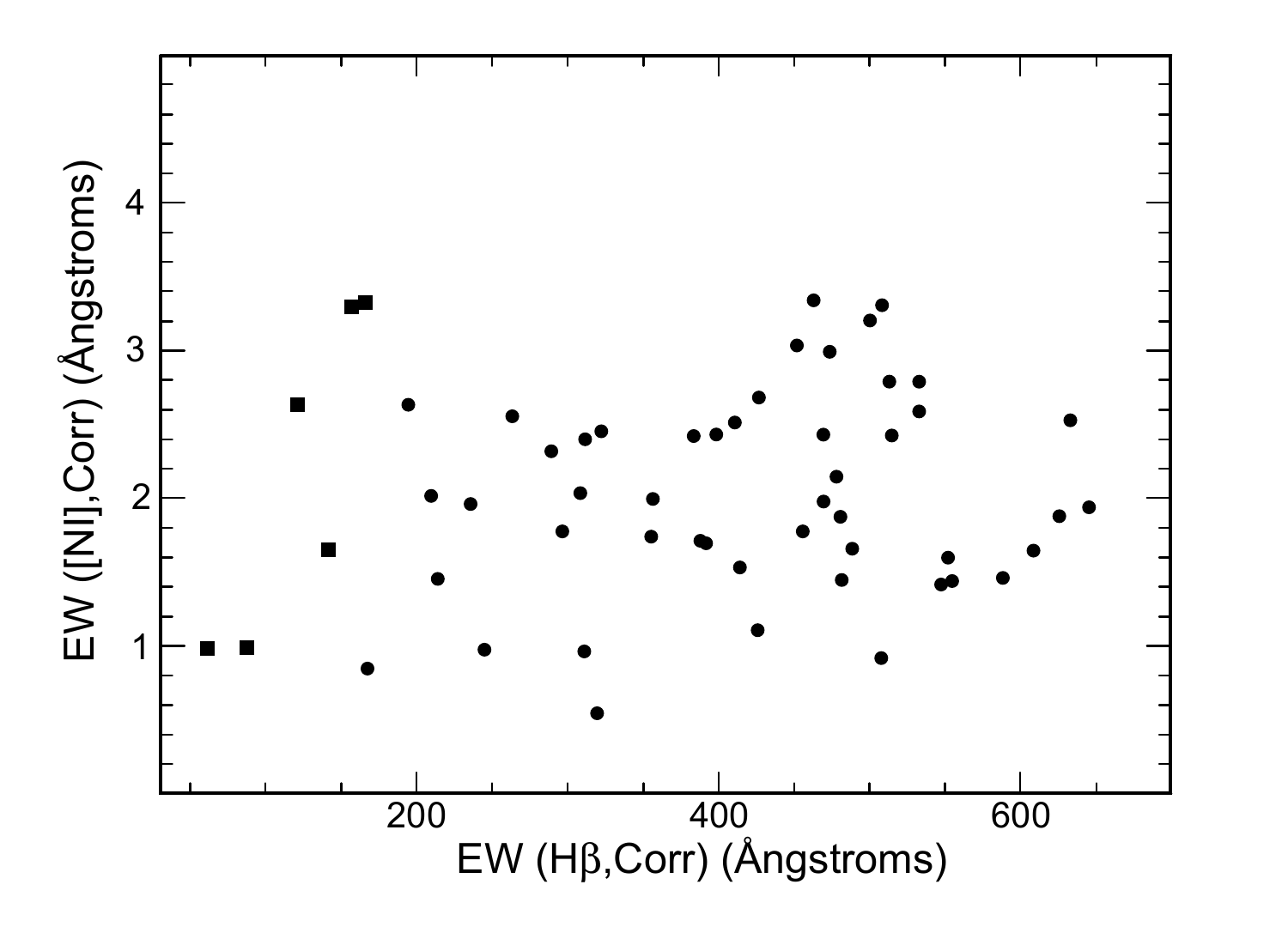}
  \caption{Equivalent width of the [\ni{}] 5199+ versus equivalent width of \Hbeta{}.}
  \label{fig:EWni}
\end{figure}

In \S~\ref{sec:emission-from-ray} we established a model of an essentially substellar point in the Orion Nebula that adequately explains the observed \(I([\ni])/I(\hb)\) line ratios.  
However, we see in Figure~\ref{fig:spectrophotometry}, Panel~A that this ratio varies across the Huygens Region and its near vicinity, rising monotonically with increasing distance from \thC.  
A similar increase is seen for the line ratios in M43 with increasing distance from NU Ori.  
On the other hand, Figure~\ref{fig:EWni} shows that the corrected equivalent width of the [\ni{}] lines is essentially constant at \(2 \pm 1~\AA\) and shows no detectable variation either within M42, nor between M42 and M43.  
Since the PDR is optically thick to the irradiating stellar continuum,
its visual scattered light is a measure of the FUV continuum that pumps
the upper states of 5199+.
The quantitative relation is determined by the scattering properties
of the grains.
These are elaborated in Appendix~\ref{sec:interpr-equiv-widths}.

Appendix~\ref{sec:interpr-equiv-widths} shows that the line intensity ratios and equivalent widths can be expressed in terms of the ratios of ``scattering'' efficiencies (albedos) and the ratios of the stellar continuum luminosities in different wavelength bands. 
The pumping contribution to 5199+ depends on the intensity of the SED around
1000\AA\ while the intensity of H$\beta$, which forms by recombination, is 
proportional to the continuum intensity at hydrogen-ionizing energies.
These are denoted by FUV and EUV in Table~\ref{tab:SEDs}.
In the simplest case we expect the 5199+/H$\beta$ intensity ratio to scale with
FUV/EUV, while the 5199+ equivalent width should scale with FUV/visual.

\subsection{The importance of the FUV / EUV ratio}

Table~\ref{tab:SEDs} lists the ratios of the average value of the spectral energy distribution (SED) \(\lambda L_{\lambda}\), calculated for the visual, FUV, and ionizing EUV bands, and for three different OB stellar populations characteristic of the inner Orion nebula (Trapezium region), the outer Orion nebula, and M43
(we will consider the Crab and Ring Nebulae, objects very different from Orion
and with strong [\ni{}] emission, in \S~\ref{sec:other_classes} below).
It can be seen from the table that the FUV/visual luminosity ratio is approximately constant between the three stellar populations in Orion (variation \(< 10\%\)), which arises because the spectral shape in all cases approximately follows the Rayleigh-Jeans behavior of \(L_{\lambda} \sim \lambda^{-2}\).
On the other hand, the relative strength of the EUV band with respect to the FUV and optical bands shows significant variation, being roughly 5 times greater for the Trapezium stars than for the exciting star of M43. 
We suggest that it is the presence or otherwise of variations in the broad band illuminating spectrum that is the principal determinant of the observed spatial variations in intensity ratios.

\begin{table}
  \caption{Spectral energy distributions over selected wavelength intervals}
  \label{tab:SEDs}
  \centering
  \begin{tabular}{l l@{\ \ }l@{\ \ }l @{\quad\ } r} \hline
             & \multicolumn{3}{c@{\quad\ }}{\dotfill{} SED ratio \dotfill} & \\ 
    Group       &  FUV/visual & EUV/visual & FUV/EUV & \EWhbi, \AA\\ \hline
    M42 inner   &  20.34 & 7.36 & \phn2.76 & 380\\
    M42 outer   &  18.18 & 4.39 & \phn4.14 & 213\\
    M43         &  19.02 & 1.40 & 13.62 & 81\\
    Ring Nebula &  76.33 & \llap{19}2.24 &  \phn0.40     &   \\
    Crab Nebula &  \phn1.10 &  1.12    &   \phn0.99    &   \\
    \hline
  \end{tabular}
  \tablecomments{Columns 2--4 show the ratio of \(\langle \lambda L_\lambda\rangle\) between different wavelength bands: EUV = 507--912~\AA; FUV = 950--1200~\AA; visual = 4800--4900~\AA\@.  Column 5 shows the ``intrinsic'' equivalent width \(L_{\hb}/L_\lambda\) of the stellar population, where the continuum luminosity \(L_\lambda\) is evaluated adjacent to the \hb{} line and line luminosity \(L_{\hb}\) is calculated as described in the text.  Results are shown for the three stellar groupings of Table~\ref{tab:stars}, using atmosphere models from \citet{2003ApJS..146..417L, Lanz.T07A-Grid-of-NLTE-Line-blanketed-Model}.  The Ring and Crab Nebulae are considered in \S~\ref{sec:other_classes}. 
 }
\end{table}

\subsection{The importance of the constant equivalent width}

The observed constant value of \EWni, coupled with the lack of variation in \(\SED\fuv/\SED\vis\) implies (equation~[\ref{eq:ratioEWni}]) that the ratio of [\ni] albedo to dust-scattering albedo is also constant within and between the nebulae: \(\albedo\opni/\albedo\dust = (4\pm2) \times 10^{-4}\). 
The Cloudy model of \S~\ref{sec:emission-from-ray} implies that \(\albedo\opni \simeq 2 \times 10^{-4}\) for a non-thermal broadening of \(5~\kmps\) if the illumination and viewing angles are both close to face-on (see Fig.~\ref{fig:hbeta}\textit{b}). 
The dust scattering effective albedo is therefore constrained to be \(0.5 \pm 0.3\), which is consistent with the expectations for back-scattering from the background molecular cloud (see Appendix~\ref{sec:albedo-scatter} and Fig.~\ref{fig:albedos-scat}) so long as the single-scattering albedo is relatively high. 

\subsection{Variation in the \(I([\ni])/I(\hb)\) ratio}\nopagebreak
The ratio \(I([\ni])/I(\hb)\) (Fig.~\ref{fig:spectrophotometry}\textit{a}) increases by roughly a factor 3 between the inner and outer regions of M42 and by a factor of 4 to 8 between M42 and M43.  
The different values of \(\SED\fuv/\SED\euv\) for the stellar populations (Table~\ref{tab:SEDs}) can fully explain the difference between M42 and M43 but can only account for half of the variation within M42 and cannot explain any of the variation within M43 (where the single dominant star means that the illuminating spectrum should be constant). 
This implies that a small systematic increase with radius within each nebula of the ratio of albedos \(\albedo\opni/\albedo_{\hb}\) may also play a role.  
The analysis of \S~\ref{sec:albedo-hb} shows that \(\albedo_{\hb} \simeq 0.1\) when the illumination and viewing angles are face-on, which, combined with the above value of \(\albedo\opni\) and using equation~(\ref{eq:ratioNIHb}), implies \(I([\ni])/I(\hb) \simeq 0.03\) for illumination by the Trapezium spectrum, as is observed for the innermost regions of M42.   
Inspection of Figure~\ref{fig:hbeta} shows that, as long as the plane-parallel approximation is valid, variations in the viewing angle cannot account for the inferred increase in \(\albedo\opni/\albedo_{\hb}\) with radius since the two albedos depend on angle in a similar way, except for close to edge-on orientations where \(\albedo\opni/\albedo_{\hb}\) is predicted to \emph{decrease}.
However, as discussed in \S~\ref{sec:breakd-infin-plane}, a finite curvature of the scattering layer has the effect of limiting the limb brightening for edge-on viewing angles, and this effect is much greater for \hb, where the scattering layer is much thicker than for [\ni].  
This effect may explain the increase in \(\albedo\opni/\albedo_{\hb}\) if the average viewing angle became increasingly edge-on in the outskirts of the nebula (see \S~\ref{sec:vari-illum-view} for further discussion). 
An alternative explanation could be an increase with radius of the turbulent broadening within the fluorescent [\ni] layer, but there is no independent evidence for such an increase. 
  


\subsection{The \(I([\ni])/I(\hb)\) ratio in other classes of nebulae}
\label{sec:other_classes}

The original motivation for this work was to calibrate models of the formation
of [\ni] lines in the relatively quiescent Orion environment, as a step towards
understanding what these lines indicate in the more exotic environments
where they are unusually strong.
Although continuum fluorescent excitation does account for the [\ni]
lines in Orion, the process cannot produce the stronger [\ni] emission
seen in planetary nebulae or the Crab supernova remnant.

The last two rows of Table~\ref{tab:SEDs} show the continuum intensity ratios
produced by the SEDs of the Ring Nebula (a $T = 1.2\times10^5 $~K Rauch
stellar atmosphere,
\citealt{ODell.C07The-Three-Dimensional-Ionization-Structure-and-Evolution})
and the Crab Nebula \citep{Davidson.K85Recent-developments-concerning-the-Crab}.
The [\ni]/H$\beta$ intensity ratio scales with the FUV/EUV continuum ratio.
Table~\ref{tab:SEDs} shows that this ratio is 7 and 3 times smaller
for the Ring and Crab Nebulae than in Orion.  
Accordingly the continuum fluorescent excitation contribution 
to the [\ni]/H$\beta$ intensity ratio produced by fluorescence will be of order 
$I([\ni])/I(\hb) \sim 10^{-3}$.
The observed line ratio is several orders of magnitude larger,
showing that other processes must be at work.
Thermal excitation by warm gas, perhaps produced by penetrating
energetic photons or ionizing particles, seems to be needed.
High resolution observations of the lines given in 
Figure~\ref{fig:varyNT} could test whether thermal processes do
account for the observed spectrum.

\section{Conclusions}
\label{sec:conclusions}

There are multiple conclusions that can be drawn from the work reported 
upon in this paper. 
Some are positive, in the sense that we find quantitative explanations 
for detailed observations of the Orion Nebula, 
while some are negative, in the sense that we demonstrate that 
the FUV pumping mechanism cannot be the dominant process 
in objects like the planetary nebula the Ring Nebula and 
supernova remnants like the Crab Nebula. 
The specific conclusions are summarized below.

\begin{enumerate}

\item  The [\ni] doublet is produced not by collisional excitation 
out of the lower-lying ground-state of neutral nitrogen. Rather, 
it is the result of FUV continuum radiation being absorbed and 
populating a higher electronic state which then populates the 
upper states of the [\ni] doublet by cascade.

\item The process operates in the thin transition boundary of the PDR that 
is close to the overlying ionization front.

\item This process means that one cannot use the relative strength of the 
two members of the [\ni] doublet as density indicators.

\item In order for this mechanism to produce the intensity of the [\ni] emission 
seen in the Huygens Region of the Orion Nebula 
there must be a non-thermal component to the broadening of the FUV absorption line that drives the process, 
with FWHM of approximately \(5~\kmps\) (see Fig.~\ref{fig:varyTurb}).
We argue in Appendix~C that the origin of this broadening 
cannot be the same as the transonic turbulence 
that is believed to be responsible for broadening the optical emission lines in the \hii{} region
because the latter operates at too large a scale 
to affect the radiative transfer in the thin pumping layer.  
Instead, we suggest that small-scale thermal instabilities may be responsible. 

\item The constant value of the equivalent width of the [\ni] doublet 
with respect to the underlying scattered light continuum 
can be interpreted as the PDR being optically thick to 
scattered starlight and a combination of reasonable assumptions 
about the scattering properties of the solid particles in the 
PDR and the orientation of the PDR.

\item  The efficacy of this pumping process is critically 
dependent upon the ratio of FUV/EUV radiation from the illuminating sources. 
We show that the stars associated with the Orion Nebula and the 
independent low ionization \hii\ region M43 explain the different 
amounts of [\ni] excess emission in these very different objects. 

\item The FUV/EUV ratio for a bright planetary nebula (the Ring Nebula) 
and the well observed Crab Nebula supernova remnant  
indicate that the FUV pumping mechanism that explains 
the Orion Nebula and M~43 is not the source of the excess [\ni] emission 
in those objects.

\end{enumerate}

Acknowledgements:  
We thank the referee for a careful review of the manuscript.
GJF acknowledges support by NSF (0908877; 1108928; and 1109061), NASA (07-ATFP07-0124, 10-ATP10-0053, and 10-ADAP10-0073), JPL (RSA No 1430426), and STScI (HST-AR-12125.01, GO-12560, and HST-GO-12309).  WJH acknowledges financial support from DGAPA-UNAM, through grant PAPIIT IN102012.
CRO was supported in part by STScI grant GO-11232.
PvH acknowledges support from the Belgian Science Policy Office through the ESA Prodex programme.
This research used data from the Atomic Line List (\url{http://www.pa.uky.edu/~peter/atomic}).

\bibliographystyle{apj}
\bibliography{new-ni-biblio}

\begin{thebibliography}{87}
\expandafter\ifx\csname natexlab\endcsname\relax\def\natexlab#1{#1}\fi

\bibitem[{{Aannestad}(1989)}]{Aannestad:1989}
{Aannestad}, P.~A. 1989, \apj, 338, 162

\bibitem[{{Abel} {et~al.}(2004){Abel}, {Brogan}, {Ferland}, {O'Dell}, {Shaw},
  \& {Troland}}]{2004ApJ...609..247A}
{Abel}, N.~P., {Brogan}, C.~L., {Ferland}, G.~J., {O'Dell}, C.~R., {Shaw}, G.,
  \& {Troland}, T.~H. 2004, \apj, 609, 247

\bibitem[{{Abel} {et~al.}(2006){Abel}, {Ferland}, {O'Dell}, {Shaw}, \&
  {Troland}}]{2006ApJ...644..344A}
{Abel}, N.~P., {Ferland}, G.~J., {O'Dell}, C.~R., {Shaw}, G., \& {Troland},
  T.~H. 2006, \apj, 644, 344

\bibitem[{{Abel} {et~al.}(2005){Abel}, {Ferland}, {Shaw}, \& {van
  Hoof}}]{Abel.N05The-H-II-Region/PDR-Connection:-Self-consistent-Calculations}
{Abel}, N.~P., {Ferland}, G.~J., {Shaw}, G., \& {van Hoof}, P.~A.~M. 2005,
  \apjs, 161, 65

\bibitem[{{Arthur} {et~al.}(2011){Arthur}, {Henney}, {Mellema}, {de Colle}, \&
  {V{\'a}zquez-Semadeni}}]{Arthur:2011}
{Arthur}, S.~J., {Henney}, W.~J., {Mellema}, G., {de Colle}, F., \&
  {V{\'a}zquez-Semadeni}, E. 2011, \mnras, 414, 1747

\bibitem[{{Arthur} {et~al.}(2004){Arthur}, {Kurtz}, {Franco}, \&
  {Albarr{\'a}n}}]{2004ApJ...608..282A}
{Arthur}, S.~J., {Kurtz}, S.~E., {Franco}, J., \& {Albarr{\'a}n}, M.~Y. 2004,
  \apj, 608, 282

\bibitem[{{Baldwin} {et~al.}(1991){Baldwin}, {Ferland}, {Martin}, {Corbin},
  {Cota}, {Peterson}, \&
  {Slettebak}}]{Baldwin.J91Physical-conditions-in-the-Orion-Nebula}
{Baldwin}, J.~A., {Ferland}, G.~J., {Martin}, P.~G., {Corbin}, M.~R., {Cota},
  S.~A., {Peterson}, B.~M., \& {Slettebak}, A. 1991, \apj, 374, 580

\bibitem[{{Baldwin} {et~al.}(2000){Baldwin}, {Verner}, {Verner}, {Ferland},
  {Martin}, {Korista}, \&
  {Rubin}}]{Baldwin.J00High-Resolution-Spectroscopy-of-Faint-Emission}
{Baldwin}, J.~A., {Verner}, E.~M., {Verner}, D.~A., {Ferland}, G.~J., {Martin},
  P.~G., {Korista}, K.~T., \& {Rubin}, R.~H. 2000, \apjs, 129, 229

\bibitem[{{Bautista}(1999)}]{Bautista:1999}
{Bautista}, M.~A. 1999, \apj, 527, 474

\bibitem[{{Berrington} \& {Burke}(1981)}]{1981P&SS...29..377B}
{Berrington}, K.~A., \& {Burke}, P.~G. 1981, \planss, 29, 377

\bibitem[{{Berrington} {et~al.}(1975){Berrington}, {Burke}, \&
  {Robb}}]{1975JPhB....8.2500B}
{Berrington}, K.~A., {Burke}, P.~G., \& {Robb}, W.~D. 1975, Journal of Physics
  B Atomic Molecular Physics, 8, 2500

\bibitem[{{Boreiko} {et~al.}(1988){Boreiko}, {Betz}, \&
  {Zmuidzinas}}]{Boreiko:1988}
{Boreiko}, R.~T., {Betz}, A.~L., \& {Zmuidzinas}, J. 1988, \apjl, 325, L47

\bibitem[{{Butler} \& {Zeippen}(1984)}]{1984A&A...141..274B}
{Butler}, K., \& {Zeippen}, C.~J. 1984, \aap, 141, 274

\bibitem[{{Cardelli} \& {Clayton}(1988)}]{1988AJ.....95..516C}
{Cardelli}, J.~A., \& {Clayton}, G.~C. 1988, \aj, 95, 516

\bibitem[{{Chandrasekhar}(1960)}]{Chandrasekhar:1960}
{Chandrasekhar}, S. 1960, {Radiative transfer} (New York: Dover)

\bibitem[{{Crutcher} {et~al.}(2010){Crutcher}, {Wandelt}, {Heiles},
  {Falgarone}, \& {Troland}}]{Crutcher:2010}
{Crutcher}, R.~M., {Wandelt}, B., {Heiles}, C., {Falgarone}, E., \& {Troland},
  T.~H. 2010, \apj, 725, 466

\bibitem[{{Davidson} \&
  {Fesen}(1985)}]{Davidson.K85Recent-developments-concerning-the-Crab}
{Davidson}, K., \& {Fesen}, R.~A. 1985, \araa, 23, 119

\bibitem[{{Dopita} {et~al.}(1976){Dopita}, {Mason}, \&
  {Robb}}]{1976ApJ...207..102D}
{Dopita}, M.~A., {Mason}, D.~J., \& {Robb}, W.~D. 1976, \apj, 207, 102

\bibitem[{{Draine}(2003)}]{Draine:2003}
{Draine}, B.~T. 2003, \apj, 598, 1017

\bibitem[{{Drawin}(1969)}]{1969ZPhy..225..483D}
{Drawin}, H.~W. 1969, Zeitschrift fur Physik, 225, 483

\bibitem[{{Ercolano} {et~al.}(2011){Ercolano}, {Dale}, {Gritschneder}, \&
  {Westmoquette}}]{Ercolano:2011a}
{Ercolano}, B., {Dale}, J.~E., {Gritschneder}, M., \& {Westmoquette}, M. 2011,
  ArXiv e-prints

\bibitem[{{Esteban} {et~al.}(2004){Esteban}, {Peimbert}, {Garc{\'{\i}}a-Rojas},
  {Ruiz}, {Peimbert}, \&
  {Rodr{\'{\i}}guez}}]{Esteban.C04A-reappraisal-of-the-chemical-composition-of-the-Orion}
{Esteban}, C., {Peimbert}, M., {Garc{\'{\i}}a-Rojas}, J., {Ruiz}, M.~T.,
  {Peimbert}, A., \& {Rodr{\'{\i}}guez}, M. 2004, \mnras, 355, 229

\bibitem[{{Esteban} {et~al.}(1999){Esteban}, {Peimbert}, \&
  {Torres-Peimbert}}]{Esteban.C99Physical-conditions-in-the-partially-ionized}
{Esteban}, C., {Peimbert}, M., \& {Torres-Peimbert}, S. 1999, \aap, 342, L37

\bibitem[{{Ferland}(1992)}]{Ferland.G92N-III-line-emission-in-planetary-nebulae}
{Ferland}, G.~J. 1992, \apjl, 389, L63

\bibitem[{{Ferland} {et~al.}(2009){Ferland}, {Fabian}, {Hatch}, {Johnstone},
  {Porter}, {van Hoof}, \&
  {Williams}}]{Ferland.G09Collisional-heating-as-the-origin-of-filament}
{Ferland}, G.~J., {Fabian}, A.~C., {Hatch}, N.~A., {Johnstone}, R.~M.,
  {Porter}, R.~L., {van Hoof}, P.~A.~M., \& {Williams}, R.~J.~R. 2009, \mnras,
  392, 1475

\bibitem[{{Ferland} \&
  {Rees}(1988)}]{Ferland.G88Radiative-equilibrium-of-high-density-clouds}
{Ferland}, G.~J., \& {Rees}, M.~J. 1988, \apj, 332, 141

\bibitem[{{Field}(1965)}]{Field:1965}
{Field}, G.~B. 1965, \apj, 142, 531

\bibitem[{{Field} {et~al.}(1969){Field}, {Goldsmith}, \& {Habing}}]{Field:1969}
{Field}, G.~B., {Goldsmith}, D.~W., \& {Habing}, H.~J. 1969, \apjl, 155, L149

\bibitem[{{Fitzpatrick} \& {Massa}(2005)}]{Fitzpatrick:2005}
{Fitzpatrick}, E.~L., \& {Massa}, D. 2005, \aj, 129, 1642

\bibitem[{{Froese Fischer} \& {Tachiev}(2004)}]{2004ADNDT..87....1F}
{Froese Fischer}, C., \& {Tachiev}, G. 2004, Atomic Data and Nuclear Data
  Tables, 87, 1

\bibitem[{{Garc{\'{\i}}a-D{\'{\i}}az} \& {Henney}(2007)}]{2007AJ....133..952G}
{Garc{\'{\i}}a-D{\'{\i}}az}, M.~T., \& {Henney}, W.~J. 2007, \aj, 133, 952

\bibitem[{{Garc{\'{\i}}a-D{\'{\i}}az}
  {et~al.}(2008){Garc{\'{\i}}a-D{\'{\i}}az}, {Henney}, {L{\'o}pez}, \&
  {Doi}}]{2008RMxAA..44..181G}
{Garc{\'{\i}}a-D{\'{\i}}az}, M.~T., {Henney}, W.~J., {L{\'o}pez}, J.~A., \&
  {Doi}, T. 2008, Revista Mexicana de Astronomia y Astrofisica, 44, 181

\bibitem[{{Godefroid} \& {Fischer}(1984)}]{1984JPhB...17..681G}
{Godefroid}, M., \& {Fischer}, C.~F. 1984, Journal of Physics B Atomic
  Molecular Physics, 17, 681

\bibitem[{{Harvey-Smith} {et~al.}(2011){Harvey-Smith}, {Madsen}, \&
  {Gaensler}}]{Harvey-Smith:2011}
{Harvey-Smith}, L., {Madsen}, G.~J., \& {Gaensler}, B.~M. 2011, ArXiv e-prints

\bibitem[{{Heiles} {et~al.}(1981){Heiles}, {Chu}, \&
  {Troland}}]{1981ApJ...247L..77H}
{Heiles}, C., {Chu}, Y.-H., \& {Troland}, T.~H. 1981, \apjl, 247, L77

\bibitem[{{Henney}(1998)}]{1998ApJ...503..760H}
{Henney}, W.~J. 1998, \apj, 503, 760

\bibitem[{{Henney}(2003)}]{2003RMxAC..15..175H}
{Henney}, W.~J. 2003, in Revista Mexicana de Astronomia y Astrofisica
  Conference Series, 175--180

\bibitem[{{Henney} {et~al.}(2005{\natexlab{a}}){Henney}, {Arthur}, \&
  {Garc{\'{\i}}a-D{\'{\i}}az}}]{2005ApJ...627..813H}
{Henney}, W.~J., {Arthur}, S.~J., \& {Garc{\'{\i}}a-D{\'{\i}}az}, M.~T.
  2005{\natexlab{a}}, \apj, 627, 813

\bibitem[{{Henney} {et~al.}(2005{\natexlab{b}}){Henney}, {Arthur}, {Williams},
  \& {Ferland}}]{2005ApJ...621..328H}
{Henney}, W.~J., {Arthur}, S.~J., {Williams}, R.~J.~R., \& {Ferland}, G.~J.
  2005{\natexlab{b}}, \apj, 621, 328

\bibitem[{{Henney} \&
  {O'Dell}(1999)}]{Henney.W99A-Keck-High-Resolution-Spectroscopic-Study}
{Henney}, W.~J., \& {O'Dell}, C.~R. 1999, \aj, 118, 2350

\bibitem[{{Hibbert} {et~al.}(1991){Hibbert}, {Biemont}, {Godefroid}, \&
  {Vaeck}}]{1991A&AS...88..505H}
{Hibbert}, A., {Biemont}, E., {Godefroid}, M., \& {Vaeck}, N. 1991, \aaps, 88,
  505

\bibitem[{{Kingdon} \&
  {Ferland}(1996)}]{Kingdon.J96Rate-Coefficients-for-Charge-Transfer}
{Kingdon}, J.~B., \& {Ferland}, G.~J. 1996, \apjs, 106, 205

\bibitem[{{Kisielius} {et~al.}(2009){Kisielius}, {Storey}, {Ferland}, \&
  {Keenan}}]{2009MNRAS.tmp..870K}
{Kisielius}, R., {Storey}, P.~J., {Ferland}, G.~J., \& {Keenan}, F.~P. 2009,
  \mnras, 870

\bibitem[{{Koyama} \& {Inutsuka}(2002)}]{Koyama:2002}
{Koyama}, H., \& {Inutsuka}, S.-i. 2002, \apjl, 564, L97

\bibitem[{{Lanz} \& {Hubeny}(2003)}]{2003ApJS..146..417L}
{Lanz}, T., \& {Hubeny}, I. 2003, \apjs, 146, 417

\bibitem[{{Lanz} \&
  {Hubeny}(2007)}]{Lanz.T07A-Grid-of-NLTE-Line-blanketed-Model}
---. 2007, \apjs, 169, 83

\bibitem[{{Le Teuff} {et~al.}(2000){Le Teuff}, {Millar}, \&
  {Markwick}}]{Le-Teuff.Y00The-UMIST-database-for-astrochemistry-1999}
{Le Teuff}, Y.~H., {Millar}, T.~J., \& {Markwick}, A.~J. 2000, \aaps, 146, 157

\bibitem[{{Lehmann} {et~al.}(2010){Lehmann}, {Vitrichenko}, {Bychkov},
  {Bychkova}, \& {Klochkova}}]{Lehmann:2010a}
{Lehmann}, H., {Vitrichenko}, E., {Bychkov}, V., {Bychkova}, L., \&
  {Klochkova}, V. 2010, \aap, 514, A34

\bibitem[{{Malkov}(1992)}]{Malkov:1992}
{Malkov}, O.~Y. 1992, Bulletin d'Information du Centre de Donnees Stellaires,
  40, 13

\bibitem[{{Mathis} {et~al.}(1981){Mathis}, {Perinotto}, {Patriarchi}, \&
  {Schiffer}}]{Mathis:1981}
{Mathis}, J.~S., {Perinotto}, M., {Patriarchi}, P., \& {Schiffer}, III, F.~H.
  1981, \apj, 249, 99

\bibitem[{{Mayers}(1962)}]{Mayers:1962}
{Mayers}, D.~F. 1962, \mnras, 123, 471

\bibitem[{{Mellema} {et~al.}(2006){Mellema}, {Arthur}, {Henney}, {Iliev}, \&
  {Shapiro}}]{2006ApJ...647..397M}
{Mellema}, G., {Arthur}, S.~J., {Henney}, W.~J., {Iliev}, I.~T., \& {Shapiro},
  P.~R. 2006, \apj, 647, 397

\bibitem[{{Moore}(1975)}]{1975stas.book.....M}
{Moore}, C.~E. 1975, {Selected tables of atomic spectra - A: Atomic energy
  levels - Second edition - B: Multiplet table; N I, N II, N III. Data derived
  from the analyses of optical spectra} (National Bureau of Standards)

\bibitem[{Moore(1993)}]{Moore.C93-Tables_spectra_H_C_N_O}
Moore, C.~E. 1993, Tables of Spectra of Hydrogen, Carbon, Nitrogen, and Oxygen
  Atoms and Ions, ed. J.~W. Gallagher (Boca Raton, FL: CRC Press), nIST
  compilation

\bibitem[{{O'Dell}(2001)}]{2001ARA&A..39...99O}
{O'Dell}, C.~R. 2001, \araa, 39, 99

\bibitem[{{O'Dell} {et~al.}(2011){O'Dell}, {Ferland}, {Porter}, \& {van
  Hoof}}]{ODell:2011}
{O'Dell}, C.~R., {Ferland}, G.~J., {Porter}, R.~L., \& {van Hoof}, P.~A.~M.
  2011, \apj, 733, 9

\bibitem[{{O'Dell} \& {Goss}(2009)}]{ODell:2009}
{O'Dell}, C.~R., \& {Goss}, W.~M. 2009, \aj, 138, 1235

\bibitem[{{O'Dell} \& {Harris}(2010)}]{ODell:2010}
{O'Dell}, C.~R., \& {Harris}, J.~A. 2010, \aj, 140, 985

\bibitem[{{O'Dell} {et~al.}(2003){O'Dell}, {Peimbert}, \&
  {Peimbert}}]{2003AJ....125.2590O}
{O'Dell}, C.~R., {Peimbert}, M., \& {Peimbert}, A. 2003, \aj, 125, 2590

\bibitem[{{O'Dell} {et~al.}(2007){O'Dell}, {Sabbadin}, \&
  {Henney}}]{ODell.C07The-Three-Dimensional-Ionization-Structure-and-Evolution}
{O'Dell}, C.~R., {Sabbadin}, F., \& {Henney}, W.~J. 2007, \aj, 134, 1679

\bibitem[{{O'Dell} \& {Yusef-Zadeh}(2000)}]{2000AJ....120..382O}
{O'Dell}, C.~R., \& {Yusef-Zadeh}, F. 2000, \aj, 120, 382

\bibitem[{{Osterbrock} \&
  {Ferland}(2006)}]{Osterbrock.D06Astrophysics-of-gaseous-nebulae-and-active}
{Osterbrock}, D.~E., \& {Ferland}, G.~J. 2006, {Astrophysics of gaseous nebulae
  and active galactic nuclei} (University Science Press)

\bibitem[{{Osterbrock} {et~al.}(1992){Osterbrock}, {Tran}, \&
  {Veilleux}}]{Osterbrock.D92Faint-emission-lines-in-the-spectrum}
{Osterbrock}, D.~E., {Tran}, H.~D., \& {Veilleux}, S. 1992, \apj, 389, 305

\bibitem[{{Patriarchi} \& {Perinotto}(1985)}]{Patriarchi:1985}
{Patriarchi}, P., \& {Perinotto}, M. 1985, \aap, 143, 35

\bibitem[{{Pequignot} {et~al.}(1991){Pequignot}, {Petitjean}, \&
  {Boisson}}]{1991A&A...251..680P}
{Pequignot}, D., {Petitjean}, P., \& {Boisson}, C. 1991, \aap, 251, 680

\bibitem[{{Rodr{\'{\i}}guez} {et~al.}(2011){Rodr{\'{\i}}guez}, {G{\'o}mez}, \&
  {Tafoya}}]{Rodriguez:2011}
{Rodr{\'{\i}}guez}, L.~F., {G{\'o}mez}, Y., \& {Tafoya}, D. 2011, \mnras, 000,
  000

\bibitem[{{R{\"o}llig} {et~al.}(2007){R{\"o}llig}, {Abel}, {Bell}, {Bensch},
  {Black}, {Ferland}, {Jonkheid}, {Kamp}, {Kaufman}, {Le Bourlot}, {Le Petit},
  {Meijerink}, {Morata}, {Ossenkopf}, {Roueff}, {Shaw}, {Spaans}, {Sternberg},
  {Stutzki}, {Thi}, {van Dishoeck}, {van Hoof}, {Viti}, \&
  {Wolfire}}]{Rollig.M07A-photon-dominated-region-code}
{R{\"o}llig}, M., {et~al.} 2007, \aap, 467, 187

\bibitem[{{Saraph} \& {Seaton}(1970)}]{1970MNRAS.148..367S}
{Saraph}, H.~E., \& {Seaton}, M.~J. 1970, \mnras, 148, 367

\bibitem[{{Schertl} {et~al.}(2003){Schertl}, {Balega}, {Preibisch}, \&
  {Weigelt}}]{Schertl:2003}
{Schertl}, D., {Balega}, Y.~Y., {Preibisch}, T., \& {Weigelt}, G. 2003, \aap,
  402, 267

\bibitem[{{Schiffer} \& {Mathis}(1974)}]{Schiffer:1974}
{Schiffer}, III, F.~H., \& {Mathis}, J.~S. 1974, \apj, 194, 597

\bibitem[{{Seaton} \& {Osterbrock}(1957)}]{1957ApJ...125...66S}
{Seaton}, M.~J., \& {Osterbrock}, D.~E. 1957, \apj, 125, 66

\bibitem[{{Shalima} {et~al.}(2006){Shalima}, {Sujatha}, {Murthy}, {Henry}, \&
  {Sahnow}}]{Shalima:2006}
{Shalima}, P., {Sujatha}, N.~V., {Murthy}, J., {Henry}, R.~C., \& {Sahnow},
  D.~J. 2006, \mnras, 367, 1686

\bibitem[{{Shaw} {et~al.}(2005){Shaw}, {Ferland}, {Abel}, {Stancil}, \& {van
  Hoof}}]{Shaw.G05Molecular-Hydrogen-in-Star-forming-Regions:}
{Shaw}, G., {Ferland}, G.~J., {Abel}, N.~P., {Stancil}, P.~C., \& {van Hoof},
  P.~A.~M. 2005, \apj, 624, 794

\bibitem[{{Sim{\'o}n-D{\'{\i}}az} {et~al.}(2011){Sim{\'o}n-D{\'{\i}}az},
  {Garc{\'{\i}}a-Rojas}, {Esteban}, {Stasi{\'n}ska}, {L{\'o}pez-S{\'a}nchez},
  \& {Morisset}}]{Simon-Diaz:2011a}
{Sim{\'o}n-D{\'{\i}}az}, S., {Garc{\'{\i}}a-Rojas}, J., {Esteban}, C.,
  {Stasi{\'n}ska}, G., {L{\'o}pez-S{\'a}nchez}, A.~R., \& {Morisset}, C. 2011,
  \aap, 530, A57

\bibitem[{{Sim{\'o}n-D{\'{\i}}az} {et~al.}(2006){Sim{\'o}n-D{\'{\i}}az},
  {Herrero}, {Esteban}, \& {Najarro}}]{2006A&A...448..351S}
{Sim{\'o}n-D{\'{\i}}az}, S., {Herrero}, A., {Esteban}, C., \& {Najarro}, F.
  2006, \aap, 448, 351

\bibitem[{{Stahl} {et~al.}(2008){Stahl}, {Wade}, {Petit}, {Stober}, \&
  {Schanne}}]{Stahl:2008}
{Stahl}, O., {Wade}, G., {Petit}, V., {Stober}, B., \& {Schanne}, L. 2008,
  \aap, 487, 323

\bibitem[{{St{\"o}rzer} \& {Hollenbach}(2000)}]{2000ApJ...539..751S}
{St{\"o}rzer}, H., \& {Hollenbach}, D. 2000, \apj, 539, 751

\bibitem[{{Tachiev} \& {Froese
  Fischer}(2002)}]{Tachiev.G02Breit-Pauli-energy-levels-and-transition}
{Tachiev}, G.~I., \& {Froese Fischer}, C. 2002, \aap, 385, 716

\bibitem[{{Tayal}(2000)}]{2000ADNDT..76..191T}
{Tayal}, S.~S. 2000, Atomic Data and Nuclear Data Tables, 76, 191

\bibitem[{{Tayal}(2006)}]{2006ApJS..163..207T}
---. 2006, \apjs, 163, 207

\bibitem[{{Tielens} \& {Hollenbach}(1985)}]{1985ApJ...291..722T}
{Tielens}, A.~G.~G.~M., \& {Hollenbach}, D. 1985, \apj, 291, 722

\bibitem[{{Weigelt} {et~al.}(1999){Weigelt}, {Balega}, {Preibisch}, {Schertl},
  {Sch{\"o}ller}, \& {Zinnecker}}]{Weigelt:1999}
{Weigelt}, G., {Balega}, Y., {Preibisch}, T., {Schertl}, D., {Sch{\"o}ller},
  M., \& {Zinnecker}, H. 1999, \aap, 347, L15

\bibitem[{{Wen} \& {O'Dell}(1995)}]{1995ApJ...438..784W}
{Wen}, Z., \& {O'Dell}, C.~R. 1995, \apj, 438, 784

\bibitem[{{Whalen} \& {Norman}(2008)}]{2008ApJ...672..287W}
{Whalen}, D.~J., \& {Norman}, M.~L. 2008, \apj, 672, 287

\bibitem[{{Williams}(2002)}]{2002MNRAS.331..693W}
{Williams}, R.~J.~R. 2002, \mnras, 331, 693

\bibitem[{{Wilson} {et~al.}(2001){Wilson}, {Muders}, {Kramer}, \&
  {Henkel}}]{Wilson:2001}
{Wilson}, T.~L., {Muders}, D., {Kramer}, C., \& {Henkel}, C. 2001, \apj, 557,
  240

\bibitem[{{Zel'Dovich} \& {Raizer}(1967)}]{ZelDovich:1967}
{Zel'Dovich}, Y.~B., \& {Raizer}, Y.~P. 1967, {Physics of shock waves and
  high-temperature hydrodynamic phenomena}, ed. R.~F. Hayes, W.D.;~Probstein
  (New York: Academic Press)

\end{thebibliography}

\renewcommand\appendixname{Appendices}
\begin{appendix}\twocolumngrid

\section{The \ni\ emission model} 
\label{sec:NIAtomicModelAppendix}

Here we describe recent improvements in the treatment of \ni\ emission 
in the spectral simulation code Cloudy.
Our model includes many emission processes because it is intended to be general,
and applicable to other environments.

\subsection{The atomic model}

\begin{figure}
  \centering
  \includegraphics{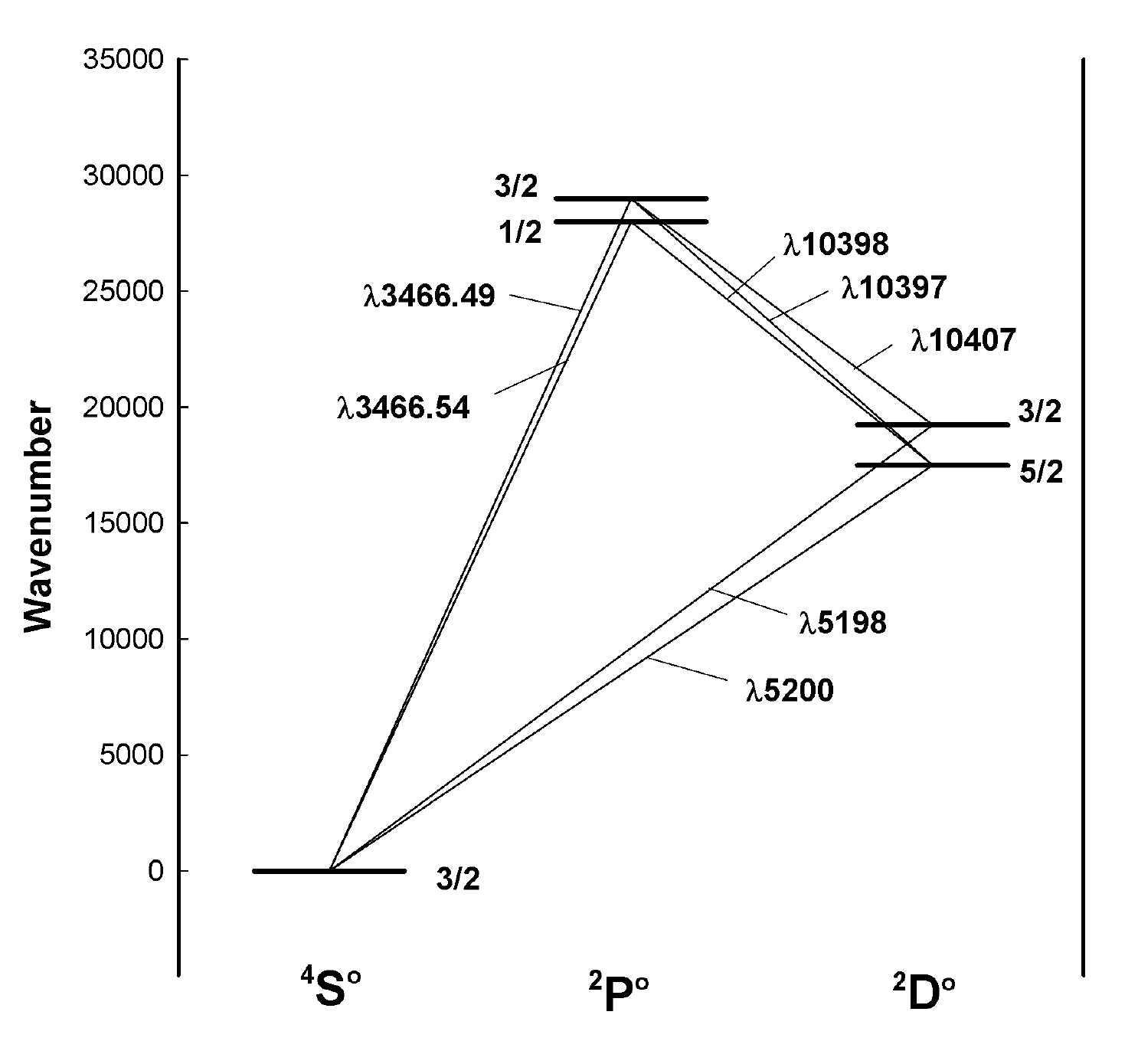}
  \caption{The lowest five levels of the \ni{} model.
  The 5198, 5200\AA\ pair of lines are denoted by $\lambda 5199^+$ in the text.
  A fourth line of the IR multiplet, $^2$P$_{1/2}$ - $^2$D$_{3/2}$ at 
  10408\AA, is not shown for clarity. }
  \label{fig:EnergyLevels}
\end{figure}

In order to optimize the speed of the model, we have chosen to model the
\ni\ atom using a 5-level atom for the metastable levels. The fluorescence
processes discussed in this paper (as well as the recombination pumping)
are added as rates populating the various excited metastable levels.
The level energies were obtained from \citet{1975stas.book.....M}
and the lowest five levels are listed in Table \ref{tab:NIEnergyLevels}
and shown in Figure~\ref{fig:EnergyLevels}.
An additional ten FUV lines can absorb photons 
in the 951\AA{}--1161\AA\ range. These lines drive the fluorescence process
and will be discussed in more detail below.

Transition probabilities have been computed by \citet{1984A&A...141..274B},
\citet{1984JPhB...17..681G}, \citet{1991A&AS...88..505H},
\citet{Tachiev.G02Breit-Pauli-energy-levels-and-transition}
and \citet{2004ADNDT..87....1F}.

\begin{table}
\centering
\caption{[\ni ] Energy Levels.}
\begin{tabular}{ c c c c c }
\hline
Configuration   &Term  &J   &Energy (cm$^{-1}$) \\
\hline
$2s^2 2p^3$     &	$^4S^o$	&3/2 &     0.000 \\
$2s^2 2p^3$     &	$^2D^o$	&5/2 &19 224.464 \\
                &           &3/2 &19 233.177 \\
$2s^2 2p^3$     &	$^2P^o$	&1/2 &28 838.920 \\
                &           &3/2 &28 839.306 \\
\hline
\label{tab:NIEnergyLevels}
\end{tabular}
\end{table}

\pagebreak[4]Table \ref{tab:NITransitionProbabilities} compares the
\citet{1984A&A...141..274B} and
\citet{1984JPhB...17..681G} rates for the forbidden transitions, referred to as the
``1984'' rates,
with the more recent calculation of \citet{2004ADNDT..87....1F},
referred to as the ``2004'' rates.
The latter rates are used. \citet{1991A&AS...88..505H} do not give
transition probabilities for the forbidden transitions. 

\begin{table}
\centering
\caption{[\ni ] transition probabilities}
\begin{tabular}{ c c c c }
\hline
air wavelength &transition  & 1984 & 2004 \\
\hline
5200.3   & $^2D^o_{5/2} \to ^4S^o_{3/2}$ & 5.77(-6)  & 7.57(-6)\\
5197.9   & $^2D^o_{3/2} \to ^4S^o_{3/2}$ & 2.26(-5)  & 2.03(-5)\\
3466.543 & $^2P^o_{1/2} \to ^4S^o_{3/2}$ & 2.52(-3)  & 2.61(-3)\\
3466.497 & $^2P^o_{3/2} \to ^4S^o_{3/2}$ & 6.21(-3)  & 6.50(-3)\\
10398.2  & $^2P^o_{1/2} \to ^2D^o_{5/2}$ & 3.03(-2)  & 3.45(-2)\\ 
10397.7  & $^2P^o_{3/2} \to ^2D^o_{5/2}$ & 5.39(-2)  & 6.14(-2)\\ 
10407.6  & $^2P^o_{1/2} \to ^2D^o_{3/2}$ & 4.63(-2)  & 5.27(-2)\\ 
10407.2  & $^2P^o_{3/2} \to ^2D^o_{3/2}$ & 2.44(-2)  & 2.75(-2)\\ 
\hline
\label{tab:NITransitionProbabilities}
\end{tabular}
\end{table}

The electron collision rates for [\ni ] have been the
subject of a number of studies.
\citet{1975JPhB....8.2500B} computed electron collision
cross sections which \citet{1976ApJ...207..102D}
converted into collision strengths.
These were later summarized by \citet{1981P&SS...29..377B}.
\citet{1976ApJ...207..102D} found some discrepancies
with existing observations and
speculated that the disagreement was due to
uncertainties in the collision strengths.
\citet{2000ADNDT..76..191T} presented close-coupling calculation of
the effective collision strengths while \citet{2006ApJS..163..207T} redid the
calculation with significantly different results.

Table \ref{tab:NICollisionStrengths} gives the history of these
electron collision strengths.
The $^4S^o  - ^2D^o$ collision strength affects the intensity
of the collisionally excited contribution to the  [\ni ] $\lambda 5199^+$ lines, while the
$^2D_{3/2}^o - ^2D_{1/2}^o$ collision strength affects the
density diagnostic.
The \citet{1981P&SS...29..377B} and \citet{2006ApJS..163..207T}
results are in reasonable agreement suggesting that
the theoretical calculations have converged onto a stable value.

\begin{table}
\centering
\caption{History of [\ni ] collision strengths at $10^4\K$}
\begin{tabular}{ c c c }
\hline
Reference  & $^4S^o  - ^2D^o$& $^2D_{3/2}^o - ^2D_{1/2}^o$ \\
\hline
Berrington \& Burke 1981&	0.48 & 0.27 \\
Tayal 2000  &	0.044        & 3.24\\
Tayal 2006  &	0.561        & 0.257\\
\hline
\label{tab:NICollisionStrengths}
\end{tabular}
\end{table}

We know of no rates for collisions with hydrogen atoms.
This should be included since we expect that [\ni] may form in 
shallow regions of the PDR, where $n(\h0) >> n_e$.
\Cloudy\ does include a general correction for H$^0$  collisions
based on the $n_e$ rate.
The effective electron density, with this correction, is
$n_e  + 1.7 \times 10^{-4} n\left( {\h0 } \right)$
based on the discussion by \citet{1969ZPhy..225..483D}.

Cloudy includes many other line formation processes in addition to thermal
collisional impact excitation 
\citep{Ferland.G88Radiative-equilibrium-of-high-density-clouds}.
Continuum pumping of the FUV lines around 1000\AA\ will be very important.
We treat this as described in 
\cite{Ferland.G92N-III-line-emission-in-planetary-nebulae} 
and \cite{Shaw.G05Molecular-Hydrogen-in-Star-forming-Regions:}.
Fluorescent excitation is mainly produced by stellar FUV photons.
Pumping will be efficient until the \ni\ lines become optically thick 
At this point they will have absorbed stellar photons over the Doppler width of the line.
Below we explore how the pumping efficiency depends on the turbulent contributor to the line width.
Other opacity sources will affect the strength of the pumped contributor to \ni\
by removing FUV photons before they are absorbed by \ni.
The two most important opacity sources are extinction of the stellar radiation field by grains within
the \hii\ region and PDR, and shielding by the forest of overlapping \htwo\ lines in deeper parts of the PDR.
These processes are all included self-consistently in our calculations.

\subsection{The fluorescence mechanism}
\label{sec:fluorescence}

In strict LS coupling, transitions that change the total spin of the atom
(called intercombination transitions) are forbidden and FUV pumping out of the
quartet ground term could not eventually populate the doublet excited terms
that produce the observed lines. However, deviations from strict LS coupling
make it possible that a significant fraction of excitations by FUV photons
will eventually populate the excited doublets. There are two possible routes:
either direct excitation by an intercombination line from the ground state or
indirect excitation by a resonance line followed by de-excitation through an
intercombination line. In the case of \ni\ both routes contribute. A complete
list of the driving lines can be found in Table~\ref{tab:NIDrivingLines}. For
each of the driving lines we calculate a 2-level atom giving us the excitation
rate for each of these transitions. We also calculated branching ratios for
the cascade down from each of the upper levels. In these calculations we
exclude the transition straight back to the ground state as this doesn't
destroy the photon. Instead it can be absorbed over and over again until
finally a different cascade from the upper level occurs (this neglects
background opacities which will be discussed further down). Intercombination
lines from the doublet system back to the quartet system are included in the
cascade, but not tracked any further after that. This implies that routes
quartet $\rightarrow$ doublet $\rightarrow$ quartet $\rightarrow$ doublet are
not included in the pumping rates. We expect the error introduced by this
approximation to be negligible. The branching ratios were calculated using
transition probabilities from \citet{2004ADNDT..87....1F}. By combining all
different routes in the cascade we could calculate a probability that an
excitation of a given driving line would result in populating any of the
metastable levels. The results of these calculations are shown in
Table~\ref{tab:NIPumpProbabilities}. The list of intercombination lines
populating the doublet metastable levels after an excitation in the {\it Ind1}
driving line is given in Table~\ref{tab:NICascade}.

\begin{table*}
\centering
\caption{The lines driving the \ni\ fluorescence. For each of the lines, the
  lower level is the ground state of \ni. The level energies are taken from
  \citet{1975stas.book.....M} and the transition probabilities from
  \citet{2004ADNDT..87....1F}. The effective collision strength is given by
  the fitting formula $\Upsilon = \exp(a + b \min[ \ln T, 10.82 ] )$ where the
  original data were obtained from \citet{2006ApJS..163..207T}.}
\begin{tabular}{lrrrrr}
\hline
label & upper level & E$_{k}$ (cm$^{-1}$) & A$_{ki}$ (s$^{-1}$) & $a$ & $b$ \\
\hline
Ind1 & $2s^2 2p^2 (3$P$) 3d$ $^4$P$_{5/2}$ & 104 825.110 & 1.62(8) & $-11.3423$ & 0.8379 \\
Dir1 & $2s^2 2p^2 (3$P$) 3d$ $^2$F$_{5/2}$ & 104 810.360 & 1.95(7) & $-12.3982$ & 0.7458 \\
Dir2 & $2s^2 2p^2 (3$P$) 3d$ $^2$D$_{5/2}$ & 105 143.710 & 8.29(5) & $ -9.4523$ & 0.3865 \\
Dir3 & $2s^2 2p^2 (3$P$) 3d$ $^2$P$_{3/2}$ & 104 615.470 & 4.29(5) & $-12.5580$ & 0.7330 \\
Dir4 & $2s^2 2p^2 (3$P$) 4s$ $^2$P$_{3/2}$ & 104 221.630 & 3.75(5) & $-10.8813$ & 0.6853 \\
Dir5 & $2s^2 2p^2 (3$P$) 3d$ $^2$P$_{1/2}$ & 104 654.030 & 2.63(5) & $-13.6532$ & 0.7712 \\
Dir6 & $2s^2 2p^2 (3$P$) 3d$ $^2$D$_{3/2}$ & 105 119.880 & 1.71(5) & $ -9.9035$ & 0.3919 \\
Dir7 & $2s^2 2p^2 (3$P$) 4s$ $^2$P$_{1/2}$ & 104 144.820 & 1.69(5) & $-11.4470$ & 0.6734 \\
Dir8 & $2s^2 2p^2 (3$P$) 3s$ $^2$P$_{3/2}$ &  86 220.510 & 4.94(4) & $ -5.4776$ & 0.1789 \\
Dir9 & $2s^2 2p^2 (3$P$) 3s$ $^2$P$_{1/2}$ &  86 137.350 & 2.72(4) & $ -6.3304$ & 0.1966 \\
\hline
\label{tab:NIDrivingLines}
\end{tabular}
\end{table*}

\begin{table*}
\centering
\caption{Probability $P_{\rm pump}$ of populating a metastable level after an
  excitation in each of the driving lines.}
\begin{tabular}{lrrrrrrrrrr}
\hline
level & Ind1 & Dir1 & Dir2 & Dir3 & Dir4 & Dir5 & Dir6 & Dir7 & Dir8 & Dir9 \\
\hline
$^2D^o_{5/2}$ & 0.0417 & 0.0468 & 0.3408 & 0.2328 & 0.7937 & 0.1338 & 0.0623 & 0.0238 & 0.6615 & 0.0000 \\
$^2D^o_{3/2}$ & 0.3441 & 0.8621 & 0.0233 & 0.0895 & 0.1068 & 0.1644 & 0.2908 & 0.8397 & 0.0694 & 0.7369 \\
$^2P^o_{1/2}$ & 0.0113 & 0.0239 & 0.0090 & 0.1617 & 0.0167 & 0.4404 & 0.4881 & 0.0876 & 0.0450 & 0.1777 \\
$^2P^o_{3/2}$ & 0.0112 & 0.0265 & 0.6253 & 0.5108 & 0.0824 & 0.2588 & 0.1569 & 0.0484 & 0.2240 & 0.0854 \\
\hline
\label{tab:NIPumpProbabilities}
\end{tabular}
\end{table*}

\begin{table}
\centering
\caption{The list of intercombination lines that can occur after an excitation
  in the {\it Ind1} driving line. For each line the upper level is $2s^2 2p^2
  (3$P$) 3d$ $^4$P$_{5/2}$. The level energies are taken from
  \citet{1975stas.book.....M} and the transition probabilities from
  \citet{2004ADNDT..87....1F}.}
\begin{tabular}{lrr}
\hline
lower level & E$_{i}$ (cm$^{-1}$) & A$_{ki}$ (s$^{-1}$) \\
\hline
$2s^2 2p^3$           $^2$D$_{3/2}$ & 19 233.177 & 1.24(7) \\
$2s^2 2p^2 (3$P$) 3p$ $^2$D$_{3/2}$ & 96 787.680 & 2.92(6) \\
$2s^2 2p^3$           $^2$D$_{5/2}$ & 19 224.464 & 1.30(6) \\
$2s^2 2p^2 (3$P$) 3p$ $^2$D$_{5/2}$ & 96 864.050 & 2.16(5) \\
$2s^2 2p^3$           $^2$P$_{3/2}$ & 28 839.306 & 1.06(5) \\
$2s^2 2p^2 (3$P$) 3p$ $^2$P$_{3/2}$ & 97 805.840 & 9.87(3) \\
\hline
\label{tab:NICascade}
\end{tabular}
\end{table}

In the previous discussion we mentioned that transitions in any of the driving
lines straight back to the ground level were not counted because these photons
would simply be re-absorbed until a different cascade occurs. This assumption
is not entirely correct as there is a finite probability $P_{\rm dest}$ that
the photon is destroyed before it can be absorbed again (e.g. due to
background opacities such as the grain opacity or bound-free opacity of
elements with sufficiently low ionization potentials). Additionally there is a
probability $P_{\rm esc}$ that the photon escapes from the cloud before it can
be absorbed again. In order to account for these processes, we modify the
excitation rate $j_2$ in s$^{-1}$ obtained from the 2-level atom as follows:
\begin{equation}
j_{\rm c} = j_2 \times \frac{1 - \beta}{1 - \beta(1 - P_{\rm dest} - P_{\rm esc})}
\end{equation}
where $\beta$ is a constant that gives the fraction of excitations in a
driving line that is followed directly by a de-excitation back to the ground
level. For {\it Ind1} $\beta = 0.7955$ and for {\it Dir1} $\beta = 0.1384$.
For all other driving lines $\beta < 0.01$ and is assumed to be zero. Given this
formula we can the write the total pump rate for each of the metastable levels as
\begin{equation}
j_{\rm meta} = \sum_i P_{\rm pump}^{\, i} \, j_{\rm c}^{\, i}
\end{equation}
where the summation runs over all the driving lines and the constants $P_{\rm
  pump}^{\, i}$ are given in Table~\ref{tab:NIPumpProbabilities} for each of the
metastable levels and driving lines.

\pagebreak[4]For completeness we should also mention that pumping of the metastable states
through recombination from N$^+$ is also included in our modeling. We use the
formulas given in \citet{1991A&A...251..680P}. These only give the rates to
the full $^2$D and $^2$P metastable terms. In our modeling we split up these
rates for each level according to statistical weight. This pumping mechanism
will of course only be effective inside the ionized region as nitrogen has a
slightly higher ionization potential than hydrogen.

From the data in Table~\ref{tab:NIDrivingLines} it is clear that all driving
lines have wavelengths longward of the Lyman limit. This implies that the
fluorescence mechanism is effective beyond the ionization front in the PDR.
Since the temperature in the PDR is generally too low to collisionally excite
the metastable doublet states, fluorescence can even become the dominant
excitation mechanism for the forbidden \ni\ lines in the PDR. It should also
be noted that even very weak direct excitation lines can have a significant
contribution to the fluorescence mechanism. If the PDR has sufficient column
density, then all driving photons will eventually be absorbed. A low
transition probability in the driving line only means that the effect is
spread over a larger area.

The fluorescence mechanism will produce permitted \ni\ emission lines that are
observable in deep spectra. The cascade routes that populate the metastable
levels will produce lines in the doublet system with wavelengths ranging
between 8567~\AA\ and 5.382~\micron{}, as well as UV lines that cannot be
observed from the ground. The shortest wavelength lines will tend to be the
strongest since they come from the lowest levels where there are only a few
alternative routes the cascade can take. In the ionized region these lines can
also be produced by recombination from N$^+$ $\to$ N$^0$, but in the PDR these
lines can {\em only} be produced by the fluorescence mechanism described here.
So if these lines are observed in the PDR, it is conclusive proof for
continuum pumping of the [\ni] lines. In Table~\ref{tab:NICascadeLines} we
list all optical cascade lines with wavelengths shorter than 1~\micron{} and in
Table~\ref{tab:NICascadeBranchProb} we list the branching probabilities for
each of the driving lines. Excitations by the {\it Dir8} and {\it Dir9}
driving lines only produce UV cascade lines and are therefore not included in
Table~\ref{tab:NICascadeBranchProb}. Each driving line has its own
characteristic spectrum. The relative strength of the contribution for each
driving line depends on the incident spectrum, the optical depth in each of
the driving lines and the escape and destruction probability for the {\it
  Ind1} and {\it Dir1} driving lines. So no generic prediction for the
spectrum can be made. However, a straight average indicates that the
$\lambda\lambda$ 9387, 9029, 9060, and 8629 lines will be the strongest, with
the $\lambda$ 9387 line having about 1.8\% of the flux of the $\lambda 5199^+$
doublet. It should be noted that the cascade lines shown in
Table~\ref{tab:NICascadeLines} are not predicted by \Cloudy.

\begin{table}
\centering
\caption{The list of permitted lines in the doublet system that can be excited
  by the fluorescence mechanism described here. Only lines with wavelengths
  between 8567~\AA\ and 1~\micron{} are listed. The transition probabilities were
  taken from \citet{2004ADNDT..87....1F}.}
\begin{tabular}{llr}
\hline
$\lambda_{\rm air}$ & transition & A$_{ki}$ (s$^{-1}$) \\
\hline
8567.735 & $2s^2 2p^2 (3$P$) 3p$ $^2$P$_{3/2}^\circ \to 2s^2 2p^2 (3$P$) 3s$ $^2$P$_{1/2}$      & 4.87(6) \\
8594.000 & $2s^2 2p^2 (3$P$) 3p$ $^2$P$_{1/2}^\circ \to 2s^2 2p^2 (3$P$) 3s$ $^2$P$_{1/2}$      & 2.10(7) \\
8629.235 & $2s^2 2p^2 (3$P$) 3p$ $^2$P$_{3/2}^\circ \to 2s^2 2p^2 (3$P$) 3s$ $^2$P$_{3/2}$      & 2.68(7) \\
8655.878 & $2s^2 2p^2 (3$P$) 3p$ $^2$P$_{1/2}^\circ \to 2s^2 2p^2 (3$P$) 3s$ $^2$P$_{3/2}$      & 1.08(7) \\
9028.922 & $2s^2 2p^2 (3$P$) 3d$ $^2$P$_{1/2}      \to 2s^2 2p^2 (3$P$) 3p$ $^2$S$_{1/2}^\circ$ & 3.20(7) \\
9060.475 & $2s^2 2p^2 (3$P$) 3d$ $^2$P$_{3/2}      \to 2s^2 2p^2 (3$P$) 3p$ $^2$S$_{1/2}^\circ$ & 3.21(7) \\
9386.805 & $2s^2 2p^2 (3$P$) 3p$ $^2$D$_{3/2}^\circ \to 2s^2 2p^2 (3$P$) 3s$ $^2$P$_{1/2}$      & 2.14(7) \\
9392.793 & $2s^2 2p^2 (3$P$) 3p$ $^2$D$_{5/2}^\circ \to 2s^2 2p^2 (3$P$) 3s$ $^2$P$_{3/2}$      & 2.52(7) \\
9395.848 & $2s^2 2p^2 (3$P$) 4s$ $^2$P$_{3/2}      \to 2s^2 2p^2 (3$P$) 3p$ $^2$S$_{1/2}^\circ$ & 1.81(4) \\
9460.676 & $2s^2 2p^2 (3$P$) 3p$ $^2$D$_{3/2}^\circ \to 2s^2 2p^2 (3$P$) 3s$ $^2$P$_{3/2}$      & 3.74(6) \\
9464.169 & $2s^2 2p^2 (3$P$) 4s$ $^2$P$_{1/2}      \to 2s^2 2p^2 (3$P$) 3p$ $^2$S$_{1/2}^\circ$ & 3.50(5) \\
\hline
\label{tab:NICascadeLines}
\end{tabular}
\end{table}

\begin{table*}
\centering
\caption{Branching probabilities of the cascade lines for each of the driving
  lines. Excitations by the {\it Dir8} and {\it Dir9} driving lines do not
  produce any optical cascade lines.}
\begin{tabular}{lrrrrrrrr}
\hline
$\lambda_{\rm air}$ & Ind1 & Dir1 & Dir2 & Dir3 & Dir4 & Dir5 & Dir6 & Dir7 \\
\hline
8567.735 & 0.0000 & 0.0001 & 0.0158 & 0.0016 & 0.0085 & 0.0015 & 0.0035 & 0.0025 \\
8594.000 & 0.0000 & 0.0000 & 0.0000 & 0.0031 & 0.0082 & 0.0129 & 0.0549 & 0.0229 \\
8629.235 & 0.0002 & 0.0008 & 0.0868 & 0.0086 & 0.0469 & 0.0082 & 0.0190 & 0.0135 \\
8655.878 & 0.0000 & 0.0000 & 0.0000 & 0.0016 & 0.0041 & 0.0065 & 0.0275 & 0.0115 \\
9028.922 & 0.0000 & 0.0000 & 0.0000 & 0.0000 & 0.0000 & 0.2840 & 0.0000 & 0.0000 \\
9060.475 & 0.0000 & 0.0000 & 0.0000 & 0.2546 & 0.0000 & 0.0000 & 0.0000 & 0.0000 \\
9386.805 & 0.0597 & 0.1270 & 0.0004 & 0.0045 & 0.0040 & 0.0196 & 0.0343 & 0.0532 \\
9392.793 & 0.0052 & 0.0058 & 0.0487 & 0.0256 & 0.0534 & 0.0000 & 0.0059 & 0.0000 \\
9395.848 & 0.0000 & 0.0000 & 0.0000 & 0.0000 & 0.0002 & 0.0000 & 0.0000 & 0.0000 \\
9460.676 & 0.0104 & 0.0222 & 0.0001 & 0.0008 & 0.0007 & 0.0034 & 0.0060 & 0.0093 \\
9464.169 & 0.0000 & 0.0000 & 0.0000 & 0.0000 & 0.0000 & 0.0000 & 0.0000 & 0.0027 \\
\hline
\label{tab:NICascadeBranchProb}
\end{tabular}
\end{table*}

At this point we should discuss the accuracy of the transition probabilities
of the intercombination lines. Accurate values for such lines are hard to
obtain since they are quite sensitive to the details of the calculation.
However, for direct excitation lines, accurate values for the transition
probability are not crucial. Using the argument from the previous paragraph it
becomes clear that an error in the transition probability would only imply
that the absorption of the driving photons would happen over a smaller or
larger area, but the total amount of pumping would remain the same when
integrated over the entire PDR. This of course assumes that the PDR is
optically thick. If that is not the case, then an error in the transition
probability would alter the escape probability of the driving line. In such
circumstances accurate transition probabilities are needed. For indirect
excitation lines, accurate transition probabilities are always needed (even
when the PDR is optically thick) since the intercombination line has to
compete with stronger, fully allowed transitions in the cascade down from the
upper level of the driving line. However, this problem is mitigated by the
fact that there is only one indirect driving line versus nine direct driving
lines. So an error in this component would only have a limited effect on the
total pumping. In Table~\ref{tab:NIAComp} we compare the transition
probabilities of the lines involved in the cascade down from the indirect
excitation using data from \citet[length form]{1991A&AS...88..505H} and
\citet{2004ADNDT..87....1F}. It is apparent that discrepancies up to 1~dex and
more can occur, indicating the difficulty in calculating these data.

\begin{table}
\centering
\caption{Comparison of the transition probability for various intercombination
  lines used in our model. For each line the upper level is $2s^2 2p^2 (3$P$)
  3d$ $^4$P$_{5/2}$. The transition probabilities (units s$^{-1}$) are taken from
  \citet[labeled ``1991'']{1991A&AS...88..505H} and \citet[labeled
    ``2004'']{2004ADNDT..87....1F}. The latter were used in our model.}
\begin{tabular}{lrr}
\hline
lower level & 1991 & 2004  \\
\hline
$2s^2 2p^3$           $^2$D$_{3/2}$ & 9.08(5) & 1.24(7) \\
$2s^2 2p^2 (3$P$) 3p$ $^2$D$_{3/2}$ & 3.43(5) & 2.92(6) \\
$2s^2 2p^3$           $^2$D$_{5/2}$ & 1.02(6) & 1.30(6) \\
$2s^2 2p^2 (3$P$) 3p$ $^2$D$_{5/2}$ & 6.48(4) & 2.16(5) \\
$2s^2 2p^3$           $^2$P$_{3/2}$ & 1.04(6) & 1.06(5) \\
$2s^2 2p^2 (3$P$) 3p$ $^2$P$_{3/2}$ & 5.67(4) & 9.87(3) \\
\hline
\label{tab:NIAComp}
\end{tabular}
\end{table}

\subsection{Density -- temperature diagnostics in the collisional excitation case}
\label{sec:NIdiagnostic}

If the lines were collisionally excited then the electron density could be determined from the
ratios of the intensities of two lines of the same ion,
emitted by different levels with nearly the same excitation energy (AGN3).
Temperature is indicated by emission from levels with different excitation energies.
Together, the gas pressure could be directly measured.
This can test whether the lines are thermally excited, and is useful
for reference by future studies which will
look into the formation of [\ni] lines in planetary nebulae, 
the Crab Nebula, and cool-core cluster filaments.

We show several emission line diagnostics for the collisionally dominated case,
using our updated atomic data and model atom.
Line pairs such as the ratio of lines of [\ni ],
\begin{equation}
R_n  = {{I({}^2D_{5/2}  \to {}^4S_{3/2} )} \mathord{\left/
 {\vphantom {{I({}^2D_{5/2}  \to {}^4S_{3/2} )} {I({}^2D_{3/2}  \to {}^4S_{3/2} )}}} \right.
 \kern-\nulldelimiterspace} {I({}^2D_{3/2}  \to {}^4S_{3/2} )}} = \lambda 5200/\lambda 5198
\end{equation}
indicate the electron density in gaseous nebulae,
as shown by \citet{1957ApJ...125...66S} and
\citet{1970MNRAS.148..367S} for [\oii].
Note that the energy order of the $J$ levels within the $^2D$ term depends on both the
charge and electronic configuration.
The line ratio is defined so that it decreases as density increases.

Every collisional excitation is followed by the emission of a photon in the low-density limit.
Since the relative excitation rates of the $^2D_{5/2}$ and $^2D_{3/2}$ levels are proportional to their
collision strengths, the ratio is
\begin{equation}
R_n (n_e \rightarrow 0 ) =
\frac{ \Upsilon(^2D_{5/2}  - ^4S_{3/2})}{\Upsilon(^2D_{3/2} - ^4S_{3/2}) } = \frac{0.337}{0.224} = 1.5
\end{equation}
This is valid when $kT \gg \delta \epsilon$, where
$\delta \epsilon$ is the difference in energies of the upper levels.
This holds for all temperatures where the optical lines emit due
to the small energy difference of the upper levels.
In the high-density limit collisional processes dominate
and set up a Boltzmann level population distribution.
The relative populations of the $^2D_{5/2}$ and $^2D_{3/2}$ levels are in the
ratio of their statistical weights, and the relative intensities
of the two lines are in the ratio
\begin{equation}
R_n \left( {n_e  \to \infty } \right) =
\frac{\omega (^2D_{5/2} )A_{\lambda 5200} }{{\omega ({}^2D_{3/2} )A_{\lambda 5198} }} = \frac{3}{2}\frac{7.57 \times 10^{ - 6} }{2.03 \times 10^{ - 5} } = 0.60
\end{equation}
The line ratio varies between these intensity limits as the density varies.
The critical density, the density where the collisional and
radiative deexcitation rates are equal, is $n_{crit} \sim 10^3~\pcc$
at $\sim 10^4~\K$.

\begin{figure}
\centering
\includegraphics{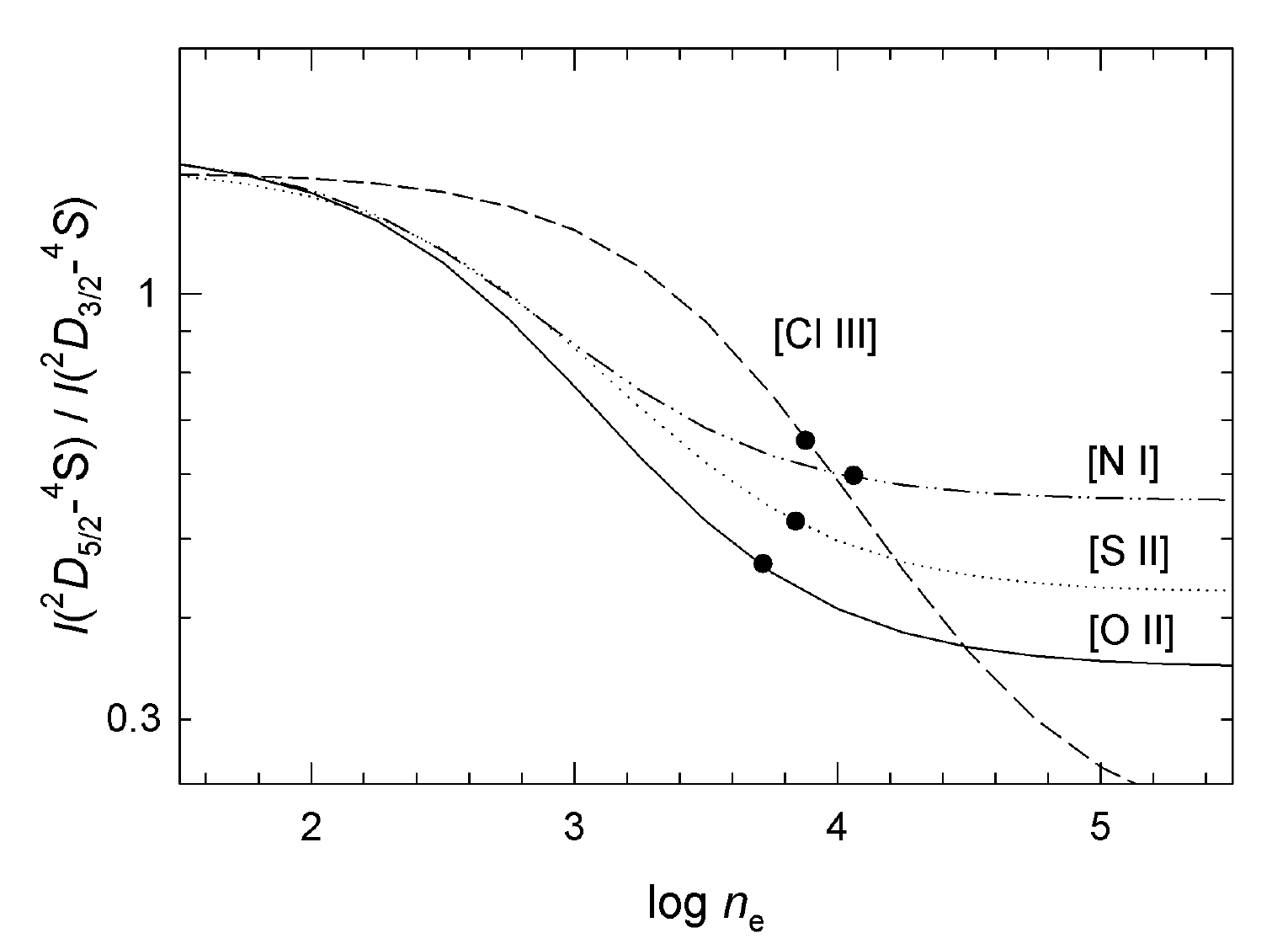}
\caption{[\ni] and three of the commonly-used density indicators present in optical spectra.
The points are from B2000.
\label{fig:DensityIndicators}
}
\end{figure}

Figure \ref{fig:DensityIndicators} compares the [\ni] $R_n$ with the
more commonly used [\oii], [\sii], and [\cliii] density indicators using data from B2000.
The [\oii] collision strengths computed by
\citet{2009MNRAS.tmp..870K} were used.
The behavior of these curves is qualitatively similar, going to the ratio of statistical
weights at low densities and a ratio that depends on the radiative
transition probabilities at high densities.

The gas kinetic temperature can be determined from ratios of intensities
from two levels with considerably different excitation energies.
For [\ni ] we have the ratio,
\begin{equation}
\begin{array}{c}
 R_T  = {{I(^2P  \to ^4S )} \mathord{\left/
 {\vphantom {{I({}^2D  \to {}^4S )} {I({}^2D \to {}^4S )}}} \right.
 \kern-\nulldelimiterspace} {I({}^2D  \to {}^4S )}} \\
  = ({\rm{3466.49  +  3466.54)/(5198  +  5200 ) = }}\lambda 3467^+/\lambda 5199^+ \\
 \end{array}
 \ .
\end{equation}

\begin{figure}
\centering
\includegraphics{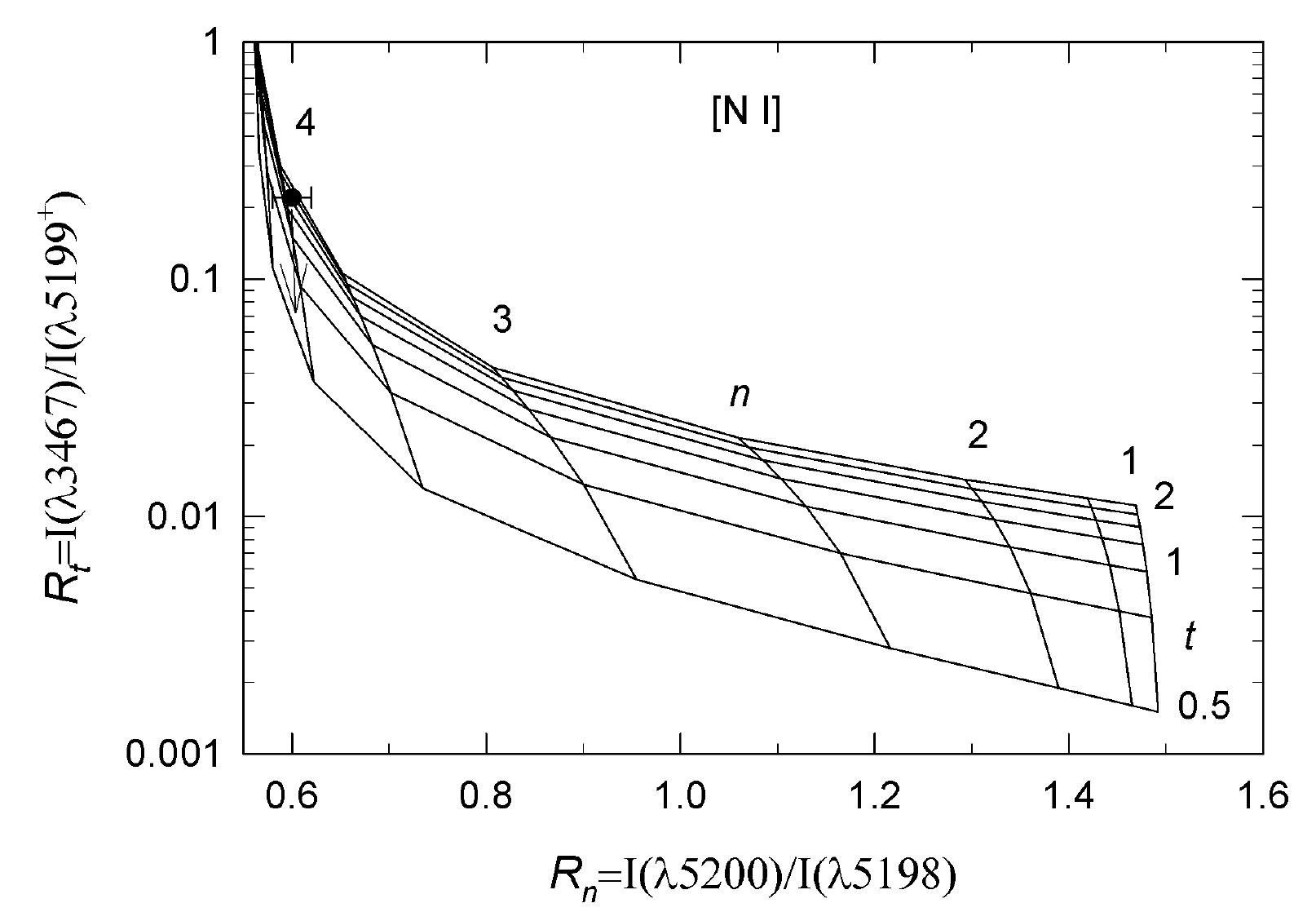}
\caption{The derived density and temperature,
in units of log n$_e$ (cm$^{-3}$) and 10$^4$ K respectively,
as deduced from line intensity ratios from
the model [\ni] atom.
\label{fig:varyNT}
}
\end{figure}
These density-temperature indicators can be combined to form a unified diagnostic diagram,
as has long been done for [\oiii ] (AGN3 Figure 5.12).
Figure \ref{fig:varyNT} shows calculated curves of the values of the
two [\ni ] intensity ratios for various
values of $T$ and $n_e$.

We used the line intensities measured in bright central regions
(B2000; 
\citealp[hereafter E2004]{Esteban.C04A-reappraisal-of-the-chemical-composition-of-the-Orion})
to estimate $n_e$ and $T$.
B2000 presented high-resolution spectrophotometric
observations of the Orion Nebula in the $\lambda\lambda 3500-7060$\AA\ range.
Their slit position was 37" west of $\theta^1$ Ori C.
This is close to the position modeled by 
\citet{Baldwin.J91Physical-conditions-in-the-Orion-Nebula}.
E2004 covered the $\lambda\lambda3100-10400$\AA\ range.
Their slit was oriented east-west and centered at
15 arcsec south and 10 arcsec west of $\theta^1$ Ori C.

\pagebreak[4]The value of $R_n$ was $0.60\pm 0.02$ for the average of the blue and
red spectra in B2000 and 0.59 for
E2004.
This is plotted in both Figures \ref{fig:DensityIndicators} and \ref{fig:varyNT}.
The results are surprising---the electron density indicated by the
[\ni] lines, which should form in partially ionized gas 
giving lower electron densities,
is 0.2--0.4~dex larger than densities indicated by lines which form
in highly ionized regions.

It is not now possible to measure the temperature using the
$\lambda3467^+$ line.
We know of no detection of this line in the Orion environment.
However \citet{Esteban.C99Physical-conditions-in-the-partially-ionized} 
reported the upper limit
of $I( 3467^+) / I(\hb ) < 10^{-3}$
from the data presented by 
\citet{Osterbrock.D92Faint-emission-lines-in-the-spectrum}.
This corresponds to $\lambda3467^+ / \lambda 5199^+ < 0.22$.

These values of the line ratios are shown in Figure \ref{fig:varyNT}.
The temperature limit indicated by the line is consistent with
formation in a photoionized environment.
The high density would be surprising if true.
Actually this can be taken as independent evidence that the lines do not form by
collisional excitation.

\subsection{Dissociation of nitrogen-bearing molecules}
\label{sec:diss-excit}

\citet{2000ApJ...539..751S} show that significant optical [\OI] emission can result from dissociation of oxygen-bearing molecules.  Could an analogous process contribute to the [\NI] emission we observe in Orion? 

\citet{2000ApJ...539..751S} consider OH photodissociation and subsequent 
[\OI] 6300+ emission in detail.  
The intensity of the line that is produced depends on the photodissociation rate, 
the branching ratio for populating the excited level producing a particular line, 
and the extinction between the point where the emission is produced 
and the surface of the cloud.

We include molecular photodissociation by the processes included in a 
modified version of the UMIST 
\citep{Le-Teuff.Y00The-UMIST-database-for-astrochemistry-1999}
database \citep{Rollig.M07A-photon-dominated-region-code}.  
Our original treatment, described in 
\citet{Abel.N05The-H-II-Region/PDR-Connection:-Self-consistent-Calculations}, considered each reaction on an ad hoc basis.  
In our upcoming release we will generalize our treatment of the chemistry to more systemically consider 
reactions, their inverses, and maintain an accounting of the consistency 
(Williams et al., in preparation).  
This is a step towards treating the chemistry as a coupled system 
that is driven by external databases.   

Using the results from the chemistry network we can identify all photodissociation processes.
The N-bearing molecules NH, CN, N$_2$, NO, and NS produce N$^0$ 
following photodissociation.  
We save this photodissociation rate per unit volume at each point in the cloud, 
and assume that each dissociation produces N$^0$ in the $^2$D$^o$ level.
Using that we can estimate the contribution of this pumping process
to the production of the $\lambda 5199^+$ lines.
The observed emission is predicted by attenuating the local emission by the 
absorption optical depth from the creation point to either side of the cloud.  
Grains are the dominant opacity source at optical wavelengths for conditions 
similar to the Orion Nebula.  
This produces an upper limit to the emission because of the assumption that 
100\% of photodissociations produce N$^0$ in the excited state producing [\NI] $5199^+$.  
This upper limit is added to the flux of the $\lambda 5199^+$ lines.

Figure \ref{fig:EmissivityVsTemperature} shows that there are regions 
of the cloud where photodissociation could make [\ni] emission.  
However this is deep enough within the cloud that the process makes no 
significant contribution to the observed flux.  
The process may be important in other environments, however.

This physics is included in the current release of \Cloudy\ (C10.00).

\section{Effective albedos for generalized scattering processes}
\label{sec:interpr-equiv-widths}

In this appendix, we develop a simple framework for evaluating the intensity of radiatively driven continuum and line processes in a photoionized nebula, which will elucidate the dependence of line ratios and equivalent widths on the shape of the exciting stellar spectrum and on geometrc factors. 
All three emission mechanisms, dust continuum, [\ni], and \Hbeta{} lines, can be thought of as diffuse reflection or scattering processes in the broadest sense, with each being driven by a different wavelength band of the stellar continuum:\footnote{The [\oi]~6300~\AA\ emission is more complicated because of a strong dependence on the ionization parameter, and so will not be considered further here.}
\begin{itemize}
\item Dust continuum is coherent scattering in the optical sense of visual band photons (\(\sim 5000~\AA\)).
\item Fluorescent [\ni] is highly incoherent scattering of FUV pumping photons (\(\sim 1000~\AA\)) into visual photons.
\item To the degree that static photoionization equilibrium holds, then \Hbeta{} emission is scattering (albeit in a statistical and indirect way) of ionizing EUV photons (\(< 912~\AA\)) into visual photons.
\end{itemize}
In each case, one can define an effective albedo \albedo{}, which is an efficiency factor that relates the intensity of scattered or emitted photons to the intensity of incident photons (see Figure~\ref{fig:scatter} and \S~\ref{sec:calc-intens-rati} below).  
Therefore, any variation in the observed intensity ratios must be due to either (1)~variations in the spectral energy distribution (SED) of the stellar radiation field, or (2)~variation in the effective albedos, or (3)~a breakdown of the simplifying assumption of a single infinite plane-parallel scattering layer.  
In the remainder of this appendix we discuss in detail the contributions of (2) and~(3), while the role of (1) is explored in \S~\ref{sec:spat-vari-intens} above.


\begin{figure}
  \centering
  \includegraphics{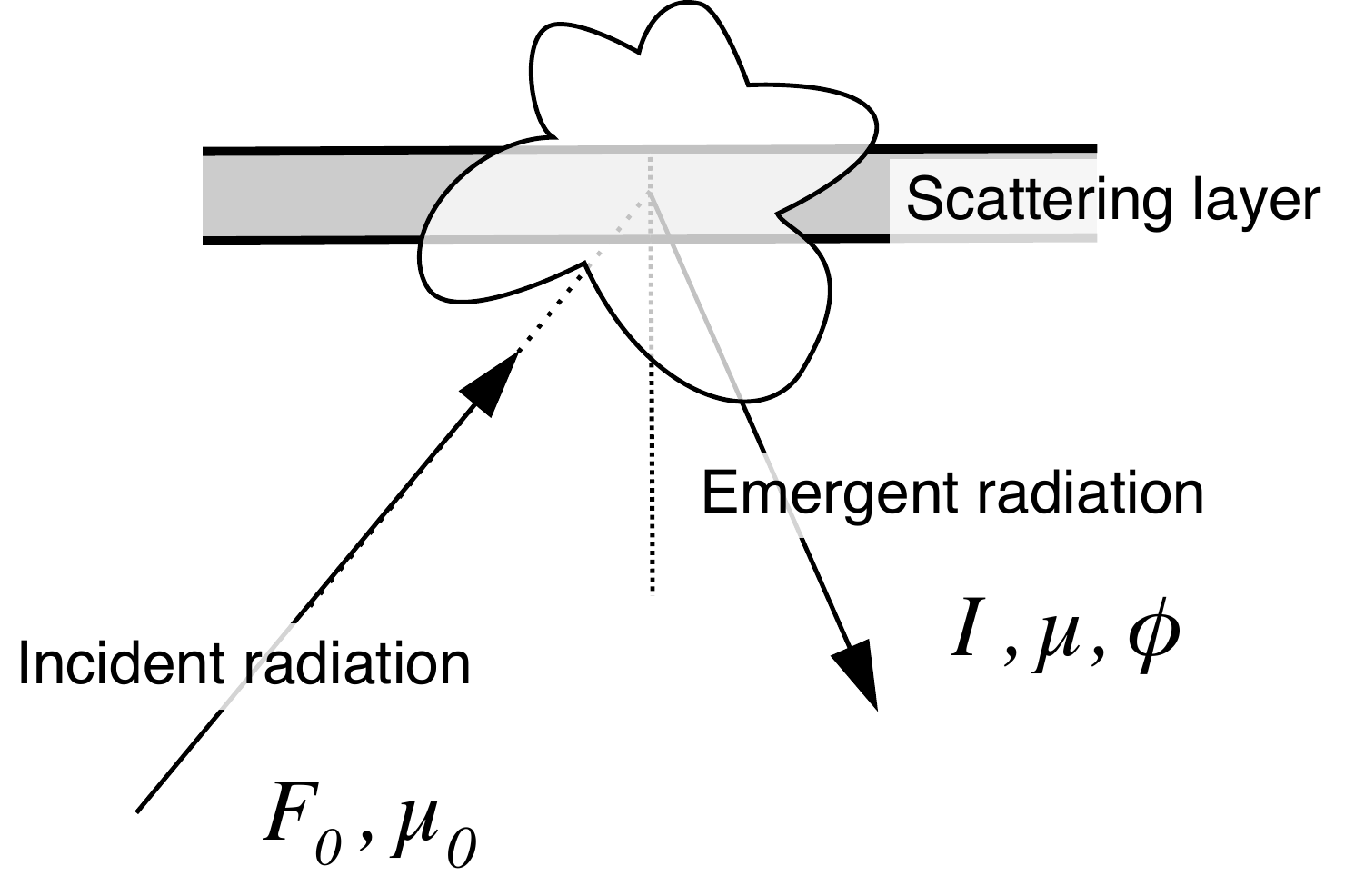}
  \caption{A black-box approach to generalized scattering processes.  
Mono-directional radiation with a flux parallel to its beam \(F_0\) (\Ufluxphot) is incident on a plane parallel scattering layer from a direction \(\mu_0 = \cos\theta_0\), where \(\theta_0\) is the angle from the normal to the layer.
The azimuth of the incident radiation may be taken as \(\phi_0 = 0\) without loss of generality.  
The intensity of emergent scattered radiation in a direction \(\mu, \phi\) is \(I\) (\Uintphot), where the scattered radiation may be in a very different wavelength band from the incident radiation.  
The effective albedo of the scattering process will depend on the directions of both the incident and emergent radiation and is defined as \(\albedo(\mu_0; \mu, \phi)  = 4 \pi I / F_0\). }
  \label{fig:scatter}
\end{figure}

\subsection{Formal calculation of intensity ratios and equivalent widths}
\label{sec:calc-intens-rati}

Under this black-box ``scattering'' or ``reprocessing'' description, the efficiency of the scattering can be described by an effective albedo \(\albedo\eff\) (see Fig.~\ref{fig:scatter}), so that the photon intensity \(I\) for each line or continuum process is proportional to the local continuum flux \(F_0\) in the spectral band that excites the scattering process: 
\begin{equation}
  \label{eq:intensity}
  I = \frac{F_0 \albedo\eff}{4 \pi} \ \Uintphot
\end{equation}
where 
\begin{equation}
  \label{eq:photonflux}
  F_0 = \frac{\SED\band (\Delta \lambda)\band}{4 \pi R^2 h c} \ \Ufluxphot. 
\end{equation}
In this expression, \(R\) is the distance from the star to the scattering layer, and \(\SED\band\) and \( (\Delta \lambda)\band\) are respectively the mean SED and wavelength width of the continuum band that excites the process.  In general, the albedo will be a function of the illumination angle and viewing angle (Fig.~\ref{fig:scatter}).  The particular values of the albedo for the production of the \hb{} recombination line, the [\ni] fluorescent lines, and dust-scattered continuum in the nebula are calculated in \S~\ref{sec:estim-effect-albed} below. 

The line ratios and equivalent widths measured in \S~\ref{sec:observations} will then be given by
\begin{equation}
  \label{eq:ratioNIHb}
  \frac{I([\ni])}{I(\hb)} = 
  \frac{\albedo\opni\, \SED\fuv\, (\Delta\lambda)\fuv}{\albedo_{\hb}\, \SED\euv\, (\Delta\lambda)\euv}
\end{equation}
\begin{equation}
  \label{eq:ratioEWHb}
  \EWhbc = 
  \frac{\albedo_{\hb}\, \SED\euv\, (\Delta\lambda)\euv}{\albedo\dust\, \SED\vis}
\end{equation}
\begin{equation}
  \label{eq:ratioEWni}
  \EWni = 
  \frac{\albedo\opni\, \SED\fuv\, (\Delta\lambda)\fuv}{\albedo\dust\, \SED\vis}
\end{equation}

A particularly simple limiting case is provided by the situation in the extreme outskirts of the Extended Orion Nebula.  
For these regions studies have shown that, except for the lowest ionization lines, essentially \emph{all} the radiation, including the emission lines, is scattered by dust rather than being emitted locally \citep{ODell:2009, ODell:2010}. 
For positions far outside the bright core of the nebula, it is reasonable to make the additional assumption that the angular distribution of the incident radiation (as seen by the scatterers) is on average similar for the continuum (which comes from the star cluster) and for the emission lines (which come from the nebular gas). 
That being the case, all geometrical factors will cancel out and the effective albedo for an emission line will be the same as that for the adjacent continuum, so that the equivalent width of the line will be simply \(\EWint = L_{\mathrm{line}} / L_\lambda\), where \(L_{\mathrm{line}}\) is the total intrinsic line luminosity of the nebula and \(L_\lambda\) is the total intrinsic continuum luminosity of the star cluster.\footnote{As described in \S~\ref{sec:observations}, the contribution to the observed equivalent widths of the atomic continuum emission from the nebular gas should first be corrected for.} 
One would therefore expect that the observed corrected equivalent widths should tend to a constant value of \(\EWint\) in the extreme outskirts of the nebula. 
Just such a behavior is seen in the observed \hb{} equivalent width (\citealp{ODell:2010}, Fig.~8), which around the outer rims of the nebulae tends to a value of \(\sim 150~\AA\) for M42 and \(\sim 100~\AA\) for M43.  
Comparison with the intrinsic \hb{} equivalent widths in Table~\ref{tab:SEDs} shows good agreement in the case of the M43, although for M42 the predicted value of \(213~\AA\) is rather higher than is observed.

In Table~\ref{tab:SEDs} we show SED ratios between different wavebands for OB stellar populations characteristic of the inner Orion nebula (Trapezium region), the outer Orion nebula, and M43.  
These can be compared with different observed emission ratios shown in Figures~\ref{fig:spectrophotometry} and~\ref{fig:EWni}: FUV/visual corresponds to \EWni; EUV/visual corresponds to \EWhbc; FUV/EUV corresponds to \(I([\ni])/I(\Hbeta)\).\footnote{Note that only 2 of these 3 quantities are independent.} 

\subsection{Estimation of effective albedos for particular processes}
\label{sec:estim-effect-albed}
\subsubsection{\hb{} recombination line}
\label{sec:albedo-hb}

Assuming a thin, plane-parallel, ionization-bounded layer of dust-free hydrogen that is illuminated by ionizing photons with a flux \(F\euv\) (\Ufluxphot) incident from a direction \(\mu_0\), then the condition of global static photoionization equilibrium is given by 
\begin{equation}
  \label{eq:ioneq}
  \mu_0 F\euv = \int \alpha\B\, n\proton n\electron \, dz, 
\end{equation}
in which \(\alpha\B\) is the ``Case B'' recombination coefficient and \(n\proton\), \(n\electron\) are the proton and electron densities. 
At the same time, the emergent intensity \(I\) (\Uintphot) of the recombination line \hb{} is given by 
\begin{equation}
  \label{eq:emerg-hb}
  I = \frac{1}{4\pi \mu} \int \alpha_{\hb} \, n\proton n\electron \, dz, 
\end{equation}
where \(\alpha_{\hb}\) is an effective recombination coefficient that only includes those recombinations that give rise to the emission of an \hb{} photon. 
Combining these, the effective albedo (see Fig.~\ref{fig:scatter}) is found to be
\begin{equation} 
  \label{eq:albedo-hb}
  \albedo_{\hb} = \mean{\!\frac{\alpha_{\hb}}{\alpha\B}\!} \, \frac{\mu_0}{\mu}, 
\end{equation}
where \(\mean{\alpha_{\hb}/\alpha\B} \simeq 0.12\) for typical \hii{} region conditions \citep{Osterbrock.D06Astrophysics-of-gaseous-nebulae-and-active}. 

\begin{figure*}[t]
  \centering
  \setlength\tabcolsep{1cm}
  \begin{tabular}{@{}ll@{}}
    (\textit{a}) & (\textit{b}) \\
    \includegraphics[height=0.3\linewidth]{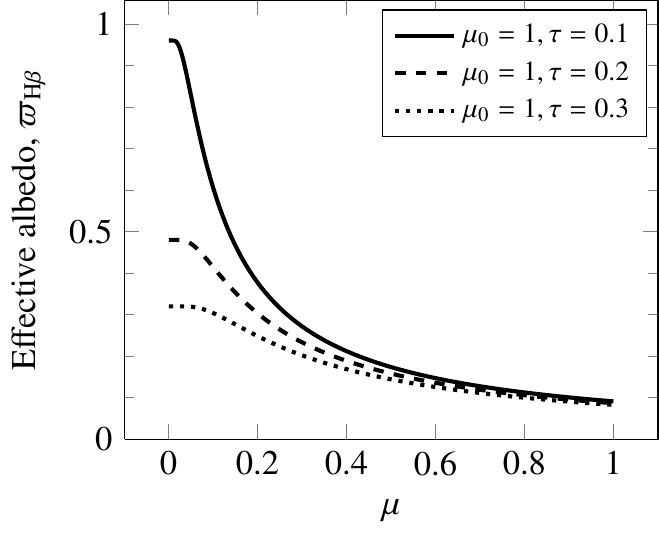} & 
    \includegraphics[height=0.3\linewidth]{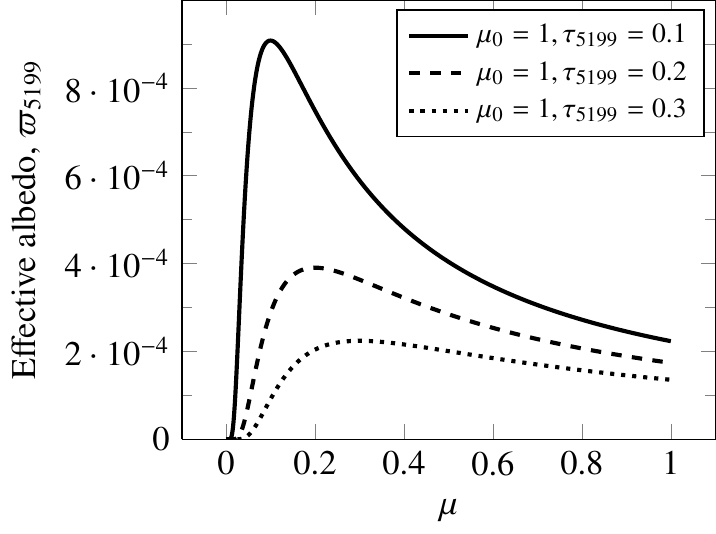}
  \end{tabular}
  \caption{(\textit{a}) Effective albedo (\(\albedo_{\hb}\), Eq.~[\ref{eq:albedo-hb-dust}]) for ``scattering'' by a plane ionized layer of normally incident ionizing EUV photons into optical \hb{} photons as a function of the viewing direction \(\mu\).  
Results are shown for three different values of \(\tau\) the perpendicular dust absorption optical depth through the layer at the wavelength of \hb{}.  
The dust acts primarily to limit the limb brightening at small values of \(\mu\).  
Note that \(\albedo_{\hb}\) is defined in terms of numbers of photons; in terms of energy the values would be a factor \( h\nu_{\hb} / \langle h\nu \rangle\euv \simeq 0.15\) times smaller. 
(\textit{b})~Same as (\textit{a}), but for fluorescent ``scattering'' of incident FUV photons into optical [\ni] photons (\(\albedo\opni\), Eq.~[\ref{eq:albedo:ni}]). }
  \label{fig:hbeta}
\end{figure*}
The correction to this result for the presence of helium will be small, but the presence of dust in the ionized gas may have a much larger effect.  
Both the incident ionizing radiation and the emergent emission line will be affected by dust absorption.
The fraction \(f\dust\) of ionizing photons that are absorbed by dust is an increasing fraction of the ionization parameter \citep{Aannestad:1989, 2004ApJ...608..282A}, but for the conditions found in Orion reaches a maximum value of about 30\% so long as the illumination is close to face-on. 
The effect is greater for edge-on illumination, but such cases, with \(\mu_0 \ll 1\), have already a small albedo and so will contribute little to the observed emission so long as a variety of illumination angles is present (see discussion in \S~\ref{sec:vari-illum-view}). 
The absorption of emergent \hb{} photons is more important since this is largest for precisely those cases \(\mu \ll 1\) which would give the highest albedo in the dust-free case (eq.~[\ref{eq:albedo-hb}]). 
If the dust absorption optical depth of the scattering layer at the wavelength of \hb{} is \(\tau\), then in the approximation that the ionized density is constant equation~(\ref{eq:albedo-hb}) becomes
\begin{equation} 
  \label{eq:albedo-hb-dust}
  \albedo_{\hb} = \left(1 - \mean{f\dust}\right) \mean{\!\frac{\alpha_{\hb}}{\alpha\B}\!} \, \frac{\mu_0}{\tau}
\left( 1 - e^{-\tau/\mu} \right). 
\end{equation}
The maximum relative boost in the albedo due to limb brightening as \(\mu \to 0\), which is infinite in the dust-free case, is now limited to \( 1/\left( 1 - e^{-\tau} \right) \), which is a factor of 3--5 for the values of \(\tau \simeq 0.1\)--\(0.3\) expected in Orion.  Note that scattering by dust of the \hb{} photons is ignored in this approximation. 

\subsubsection{Fluorescent {\upshape [\ni]}}
\label{sec:albedo-ni}

The FUV fluorescent pumping of the optical [\ni] lines is only efficient at wavelengths where the opacity of the pumping line exceeds the background continuum opacity, which at FUV wavelengths is dominated by dust.  This gives a limit \(\delta \lambda = \lambda \delta v / c\) to the wavelength interval that contributes to the pumping, where \(\delta v\) is of order the Doppler width of the line.\footnote{
  For a Gaussian line profile, \(\delta v =  b \sqrt{\ln(k_0/k\dust)} \), where \(b = 0.601 \times \mathrm{FWHM}\) is the Doppler broadening parameter, \(k_0\) is opacity at line center, and \(k\dust\) is the continuum dust opacity.}  If there are a number \(\Nline\) pumping lines, each of effective width \(\delta v\), then the fraction of the total FUV continuum that contributes to the pumping is \(\simeq \Nline (\delta v / c) (\langle\lambda\rangle\fuv / (\Delta\lambda)\fuv)\), where \(\langle\lambda\rangle\fuv\) is the average wavelength of the pumping lines and \((\Delta\lambda)\fuv\) is the wavelength width of the FUV band.\footnote{For the wavelength range of 950--1200~\AA{} used in Table~\ref{tab:SEDs}, \(\langle\lambda\rangle\fuv / (\Delta\lambda)\fuv = 4.3\).} 
If a fraction \(f\opni\) of all pumps results in the emission of a line in the optical [\ni] \(\lambda\lambda 5198,5200~\AA\) doublet, then the effective albedo for ``scattering'' of FUV continuum into these lines is
\begin{equation}
  \label{eq:albedo:ni}
  \begin{split}
    \albedo\opni & = \frac{ 4\pi I\opni }{ F\fuv } \\
    & = f\opni \Nline
    \left(\frac{\delta v}{c}\right)
    \left(\frac{\langle\lambda\rangle\fuv}{(\Delta\lambda)\fuv}\right)
    \left(\frac{\mu_0}{\mu}\right)\, e^{-\tau\fuv/\mu_0} \,
    e^{-\tau\opni/\mu},
  \end{split}
\end{equation}
where \(\tau\fuv\) and \(\tau\opni\) are the continuum absorption optical depths between the star and the pumping layer, measured perpendicular to the layer, and at the wavelengths of the pumping FUV lines and the emerging optical lines, respectively.  

The opacity of the FUV pumping lines \(\tau\pump\) is proportional to the abundance of \nzero, and so is very low inside the ionized gas, rising suddenly at the ionization front. 
Therefore, for strong pumping lines, the fluorescent excitation (which peaks at \(\tau\pump \simeq \mu_0\)), is concentrated in a thin layer just behind the ionization front so that \(\tau\fuv\) and \(\tau\opni\) are insensitive to variations in the illumination cosine \(\mu_0\). 
For the weakest pumping lines, on the other hand, the pumping layer extends deeper into the neutral PDR and so \(\tau\fuv\) and \(\tau\opni\) are generally larger and become roughly proportional to \(\mu_0\). 

Figure~\ref{fig:hbeta}\textit{b} shows results for \(\albedo\opni\) for the case of perpendicular illumination \(\mu_0 = 1\) and assuming \(\Nline = 10\), \(\delta v = 10~\kmps\), \(f\opni = 0.1\), \(\langle\lambda\rangle\fuv / (\Delta\lambda)\fuv = 4.3\), and \(\tau\fuv = 1.5 \tau\opni\).  
It can be seen that the same optical depth of dust has a considerably larger effect on the [\ni] albedo than on the \hb{} albedo, particularly for oblique viewing angles (small \(\mu\)). 
This is because the dust absorption layer completely overlies the fluorescent scattering layer in the [\ni] case, whereas in the case of \hb{} the dust is mixed in with the line-emitting gas.
As a result, whereas the \hb{} albedo \(\albedo_{\hb}\) simply saturates at small \(\mu\), the [\ni] albedo \(\albedo\opni\) has a maximum at \(\mu = \tau\opni\) and then drops to zero as \(\mu \to 0\).

\subsubsection{Scattered starlight}
\label{sec:albedo-scatter}

The dust scattering of starlight in the nebula can be divided into two parts: (1)~back-scattering by dust in the PDR and molecular cloud located behind the the nebula, which has a high optical depth, and (2)~small-angle scattering by dust located in the diffuse clouds in front of the nebula (the neutral veil, \citealp{2004ApJ...609..247A, 2006ApJ...644..344A}), which has a smaller optical depth (\(\tau = 0.1\)--\(1\), \citealp{2000AJ....120..382O}). 
In both cases, the results will be sensitive to the optical properties of the dust grains, which at the simplest level can be characterized by the single-scattering albedo \(\albedo_0\), which is the probability that a photon interacting with a grain is scattered rather than absorbed, and the asymmetry parameter \(g\), which is the mean cosine of the scattering angle (\(g = 0\) for isotropic scattering). 
Dust in Orion is found to have a high value of the total/selective extinction ratio \(R_V \simeq 5\), possibly due to grain coagulation \citep{1988AJ.....95..516C}.  
Theoretical calculations of the optical properties of a grain population with this value of \(R_V\) (Fig.~4 of \citealp{Draine:2003}) imply that at optical wavelengths (\(\sim 5000~\AA\)) the albedo is relatively high (\(\albedo_0 \simeq 0.8\)) and the scattering is moderately forward-throwing (\(g \simeq 0.6\)), whereas at FUV wavelengths (\(\sim 1000~\AA\)) the albedo is lower (\(\albedo_0 \simeq 0.4\)) and the scattering is extremely forward-throwing (\(g \simeq 0.8\)).
Observations in Orion of scattered FUV continuum \citep{Shalima:2006} and scattered optical emission lines (\S~3.1 of \citealp{1998ApJ...503..760H}) are consistent with these values, although in both cases it is only a combination of \(\albedo_0\) and \(g\) that is constrained.  
Earlier studies \citep{Schiffer:1974, Mathis:1981, Patriarchi:1985} have found different values, and even evidence that the dust properties vary with position, but the results are very sensitive to the assumed geometry of the scattering. 
In the following we present results for both high-albedo and low-albedo grains, which can be taken as representative of the range of possible optical grain properties at visual and FUV wavelengths. 

\begin{figure*}
  \centering
  \setlength\tabcolsep{1cm}
  \begin{tabular}{ll}
    (\textit{a}) & (\textit{b}) \\
    \includegraphics[height=0.3\linewidth]{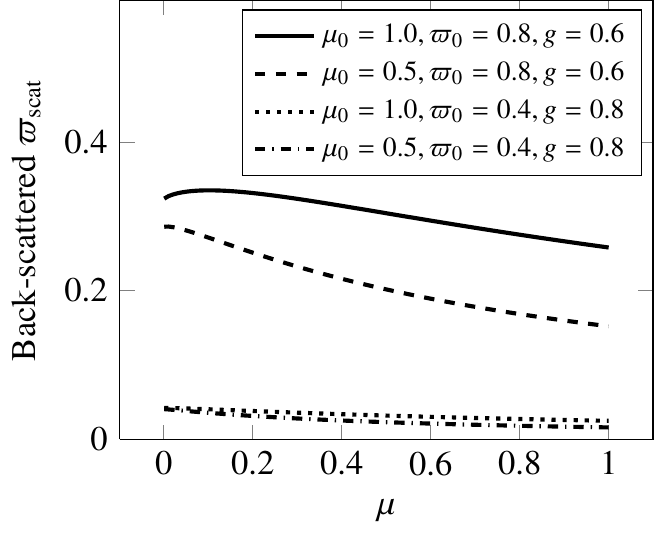} &
    \includegraphics[height=0.3\linewidth]{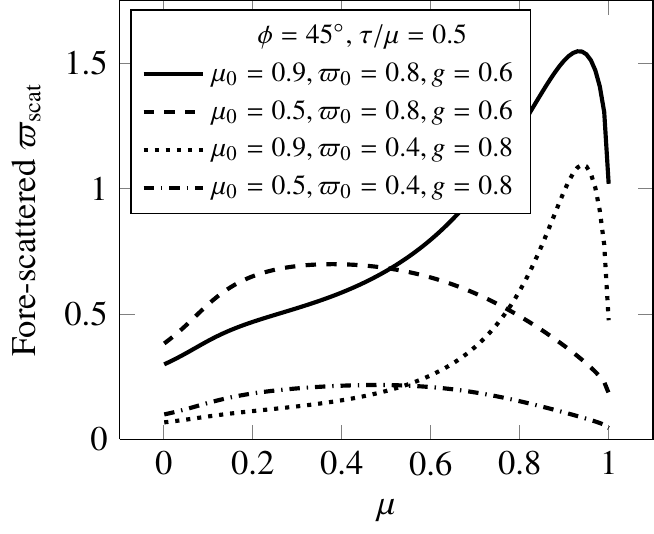}
  \end{tabular}
  \caption[]{(\textit{a})~Effective albedos for back-scattering from a very optically thick dusty layer as a function of the viewing angle \(\mu\).  
The upper two lines (solid and dashed) are for optical properties typical of Orion dust at visual wavelengths, while the lower two lines (dotted and dot-dashed) are for those of the same dust at FUV wavelengths, in both cases for two different illumination angles \(\mu_0\).  
Note that the approximation used in deriving these curves becomes less accurate as \(\mu_0\) decreases, as the results start to develop an additional dependency on the viewing azimuth \(\phi\).
(\textit{b})~Effective albedos for diffuse transmission through a translucent foreground layer with optical depth 0.5 along the line of sight. 
Line types as in (\textit{a}).}
  \label{fig:albedos-scat}
\end{figure*}

The problem of back-scattering by the molecular cloud has a well-known solution in the case of isotropic scattering \citep{Chandrasekhar:1960}, giving an effective albedo of
\begin{equation}
  \label{eq:albedo-scat}
  \albedo\dust = \albedo_0 \,\frac{\mu_0}{\mu + \mu_0}\, H(\mu)\, H(\mu_0)
\end{equation}
where \(H(\mu)\) is the Chandrasekhar H-function.
An approximate analytic form for the H-function  \citep{1998ApJ...503..760H}, accurate to \(< 5\%\) for \(\albedo_0 \le 0.9\) is 
\begin{equation}
  \label{eq:chandra-h} 
  H(\mu) = 1 + 0.5 \albedo_0\, \mu \left( 1 + 1.8 \mu^{0.4} \albedo_0^2 \right) \ln \left(1 + \mu^{-1}\right) .
\end{equation}
For the more relevant case of asymmetric scattering (\(g \ne 0\)), the problem is more difficult to solve, and is no longer axially symmetric unless \(\mu_0 = 1\). 
However, for illumination angles that are not far from face-on, a good approximation is found by simply multiplying the isotropic results by a factor of \((1 - g)^{3/2}\) \citep{1998ApJ...503..760H}.
The results of this approximation are shown in Figure~\ref{fig:albedos-scat}\textit{a}, where it is seen that typical values of \(\albedo\dust = 0.2\)--\(0.3\) are obtained at visual wavelengths, but much smaller values (\(\albedo\dust < 0.05\)) are seen at FUV wavelengths. 
In both cases, the scattering is approximately Lambertian (brightness independent of viewing angle) when the illumination is close to face on. 
Note however, that this approximation ignores the fact that as \(\mu_0\) is decreased, then the forward-throwing part of the phase function begins to be sampled at small \(\mu\) for favorable viewing azimuths \(\phi\), which would tend to increase the limb brightening for \(\mu_0 < 1\). 

For the case of forward scattering, one can use the results for diffuse transmission through a homogeneous plane parallel layer of optical thickness \(\tau\) \citep{Chandrasekhar:1960}, where the effective albedo can be expressed in terms of Chandrasekhar's X and Y functions: 
\begin{multline}
  \label{eq:albedo-forward}
  \albedo\dust = \mu_0 \, (\mu - \mu_0)^{-1} \albedo_0\, \Phi(\mu,
  \mu_0, \phi) \\
  \times \left[
    Y(\mu, \tau) X(\mu_0, \tau) - X(\mu, \tau) Y(\mu_0, \tau)
  \right]. 
\end{multline}
where \(\Phi\) is the scattering phase function and \(X\) and \(Y\) depend implicitly on \(\Phi\) and \(\albedo_0\).  
In the limit of small \(\tau\), it is sufficient to include only single scattering, which yields the approximation \(X^{(1)}(\mu, \tau) = 1\), \(Y^{(1)}(\mu, \tau) = e^{-\tau/\mu}\).
For multiple scattering in the isotropic case, extensive tables have been published for \(X\) and \(Y\) (e.g., \citealp{Mayers:1962}) and we find that an acceptable approximation to these results is given by 
\begin{equation}
  X(\mu, \tau) \simeq 1 + 0.75 A, \quad Y(\mu, \tau) \simeq (1 + 1.5 A)\, e^{-\tau/\mu}
  \label{eq:XY}
\end{equation}
where 
\begin{equation}
A = \albedo_0^2 \,\frac{2 \tau}{1 + \tau}\, \mu^{\tau/(1+\tau)}. 
\end{equation}
This approximation is accurate to \(< 10\%\) for all the cases covered by \citet{Mayers:1962} (\(\tau = 0.1\)--\(5\), \(\albedo_0 = 0.5\)--\(1\)).
To extend this result to forward-throwing phase functions, we assume that the anisotropy can be neglected for all orders of scattering higher than the first, so that equations~(\ref{eq:XY}) and (\ref{eq:albedo-forward}) may be directly combined. 

Example results for scattering from foreground dust in this approximation are shown in Figure~\ref{fig:albedos-scat}\textit{b}, assuming \(\phi = 45\arcdeg\) and a Henyey-Greenstein form for the scattering phase function: 
\begin{equation}
  \label{eq:HG}
  \Phi(\mu, \mu_0, \phi) = \frac{1-g^2}{(1 + g^2 - 2 g \mu\scat)^{3/2}}
\end{equation}
where
\begin{equation}
  \mu\scat = \mu\mu_0 + (1-\mu^2)^{1/2} (1-\mu_0^2)^{1/2} \cos\phi . 
\end{equation}
The results are shown for a fixed value of the optical depth measured along the line of sight, \(\tau/\mu\), since it is this quantity that is constrained by observations of the extinction in the neutral veil.  
Therefore, the actual thickness of the layer \(\tau\) goes to zero as \(\mu \to 0\) and no limb brightening is seen.  
Instead, the albedo tends to have a maximum when \(\mu \simeq \mu_0\) since this maximises \(\Phi\) for the small values of \(\phi\) considered here.  
It can be seen that the effective albedo is generally of order \(\albedo_0 \tau/\mu\), although it can be several times larger than this for favorable combinations of \(\mu\), \(\mu_0\), and \(\phi\) that give sufficiently small scattering angles (\(\mu\scat > 0.8\) for the optical-band grain properties, or \(\mu\scat > 0.9\) for the FUV-band grain properties).

\subsection{Variations within the nebula of the illumination and viewing angles}
\label{sec:vari-illum-view}

\begin{figure}
  \centering
  \setkeys{Gin}{width=\linewidth}
  \begin{minipage}{0.9\linewidth}
    (a)\\
    \includegraphics{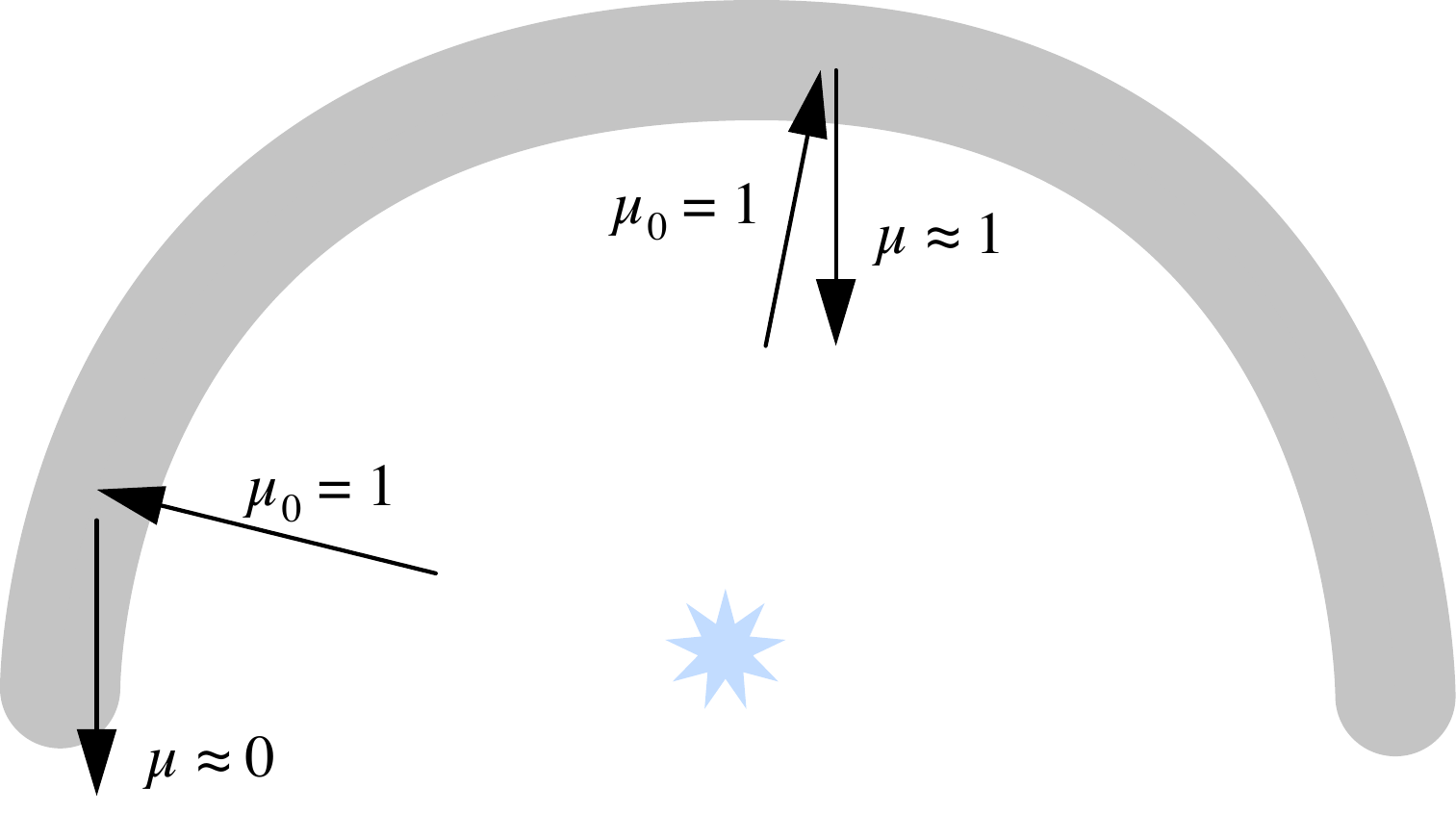}\par
    \bigskip \bigskip \bigskip
    (b)\\
    \includegraphics{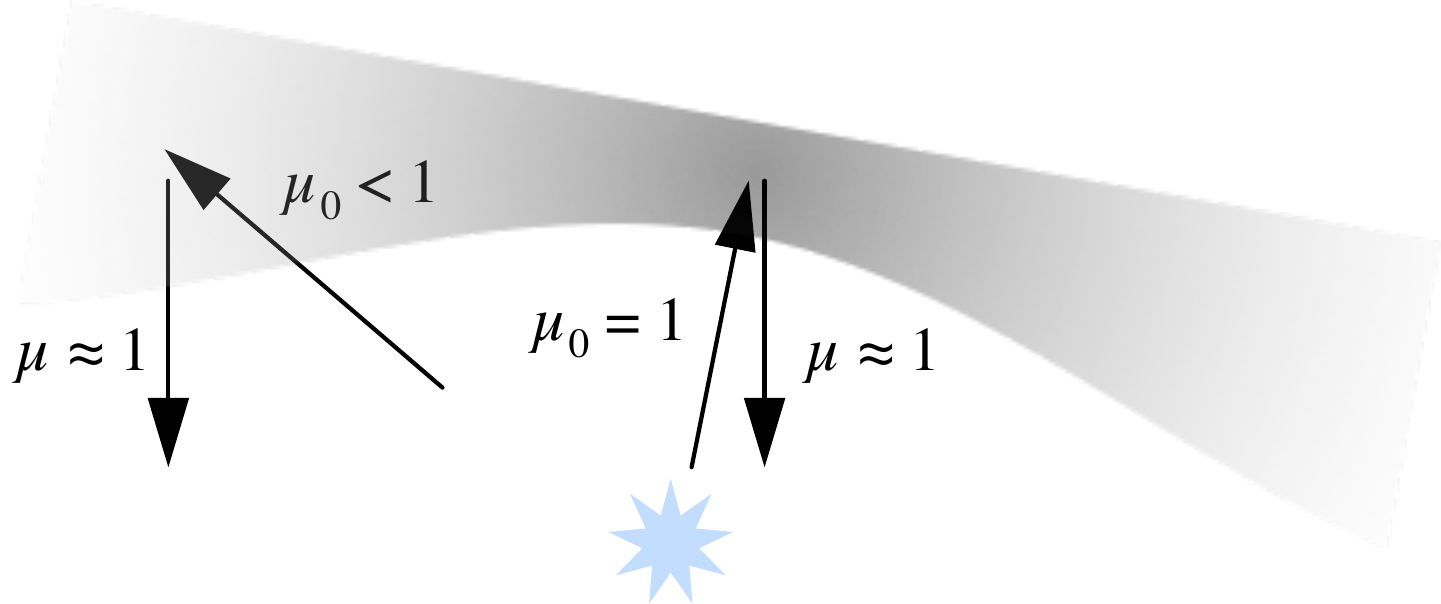}\par
    \bigskip \bigskip \bigskip
    (c)\\
    \includegraphics{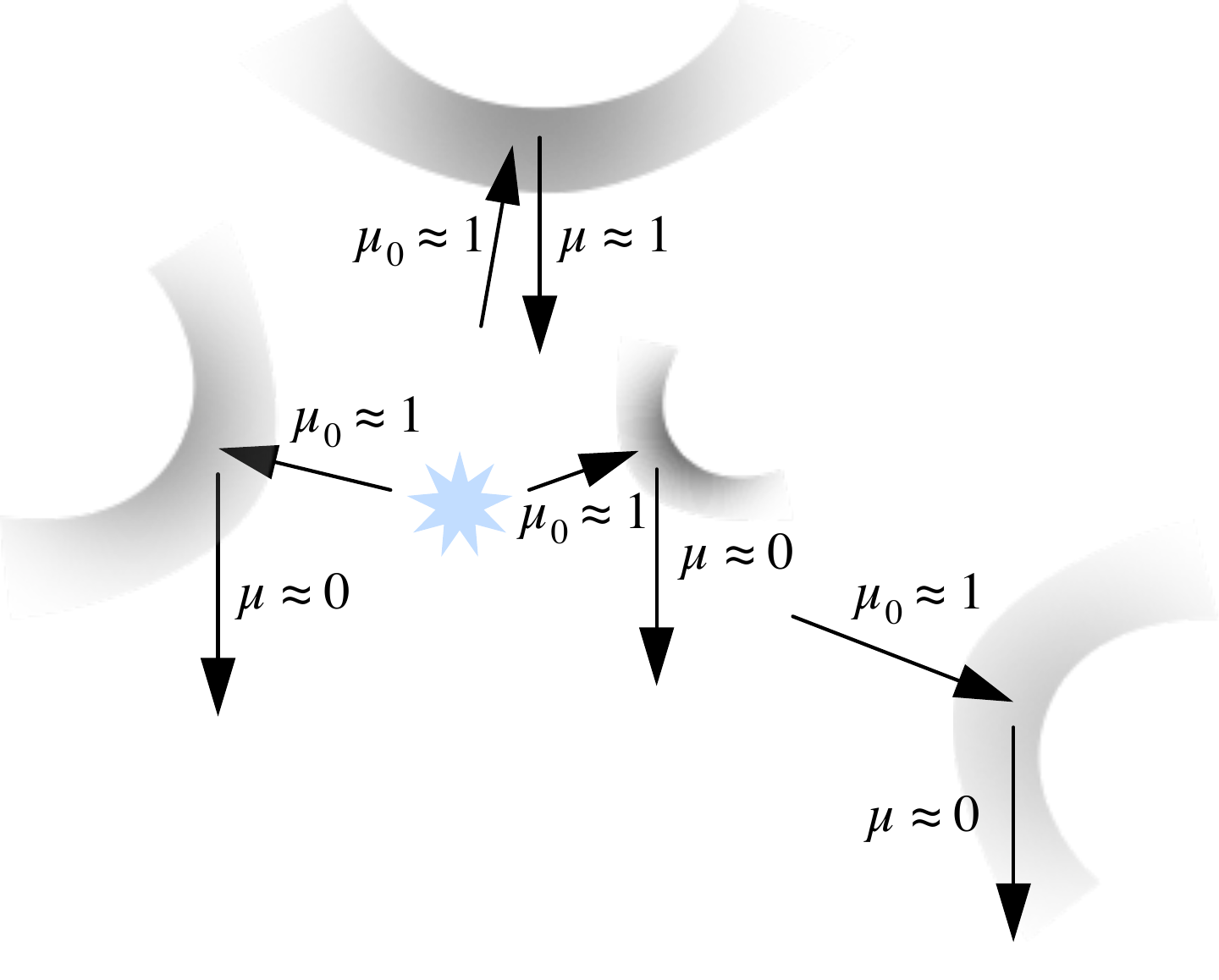}
  \end{minipage}
  \caption{Three simple models for the geometry of the nebula, showing how the illumination angle \(\mu_0\) and viewing angle \(\mu\) vary with position in the nebula.  In each case, the observer is located off the page to the bottom.  (\textit{a})~A hemispherical shell.  (\textit{b})~A nearly plane layer. (\textit{c})~An irregular nebula consisting of many globule-like and bar-like features.}
  \label{fig:geometry}
\end{figure}

The effective scattering albedos derived in the previous sections are strong functions of the angle of illumination of the scattering layer \(\mu_0\) and of the observer's viewing angle \(\mu\), with the albedo generally being highest when the illumination is close to face-on (\(\mu_0 \simeq 1\)) and the view is close to edge-on (\(\mu \simeq 0\)).  
It is therefore important to consider whether these angles vary systematically between the core and the outskirts of the nebula.  
This depends critically on the large-scale geometry of the scattering layers within the nebula.  
For instance, if the nebula were a simple hemispherical shell centered on the Trapezium stars (illustrated in Fig.~\ref{fig:geometry}\textit{a}), then the illumination angle would be constant at \(\mu_0 = 1\) while the viewing angle would vary from \(\mu = 1\) in the center to \(\mu = 0\) at the edge.  
On the other hand, if the nebula were a plane parallel layer (as in the models of \citealp{2005ApJ...627..813H} and illustrated in Fig.~\ref{fig:geometry}\textit{b}), then \(\mu\) would be constant, whereas \(\mu_0\) would vary from \(\simeq 1\) at the center to \(\simeq 0\) towards the edges. 
In reality, neither of these simple geometries works well as model for the nebula. 
In particular, the hemispherical-shell model would predict a constant ionized gas density and a surface brightness that \emph{increases} with radius, both in violent disagreement with observations. 
On the other hand, the plane-layer model fails to explain the fine-scale structure see in many emission lines (e.g., \citealp{2000AJ....120..382O, 2007AJ....133..952G}), as well as the sharp edge of the EON\@. 
For an observational aperture that is larger than the angular size of the individual emission structures, the observed emission will be biased towards face-on illumination angles and edge-on viewing angles, simply because those are the cases that give the highest effective albedo. 


\subsection{Breakdown of the infinite plane-parallel layer approximation}
\label{sec:breakd-infin-plane}

The results of the previous sections assume that the scattering occurs in a single plane-parallel layer of infinite lateral extent, which is a good approximation so long as the thickness of the each scattering layer and the displacements between them are much smaller than either the distance from the illuminating source, or radius of curvature of the layer, or the size of the observational aperture.  
Obviously, these conditions will be violated to a greater or lesser extent in a real nebula, which will lead to a variety of aditional effects on the line ratios. 
The most important of these can be characterized as (1)~\textit{ionization stratification}, (2)~\textit{differential pre-attenuation}, or (3)~\textit{limb-brightening limiting}. 
We now discuss these in turn and show that none of them is likely to have an important effect on the observational results discussed in this paper. 

\pagebreak[4]Ionization stratification is the angular separation on the plane of the sky of the different scattering layers, such as the separation of the \hb{} emission, which arises in the ionized gas, from the dust-scattered optical continuum, which arises predominantly in the neutral PDR\@.  
This stratification is not visible in a true plane-parallel geometry unless the viewing angle is strictly edge-on (\(\mu = 0\)), but for a finite geometry it will occur for \(|\mu| \lesssim z/R\) where \(z\) is the separation between the layers and \(R\) is the smaller of the radius of curvature or the lateral extent of the layers.  
Although ionization stratification will produce fine-scale variations in the line ratios and equivalent widths, it will not affect the values given in Figure~\ref{fig:spectrophotometry} unless the angular size corresponding to the inter-layer separation \(z\) is larger than the size of the observational sample regions. 
The sizes of the sample regions are listed in Appendix~A of \citet{ODell:2010} and range from about 1 to 7 arcminutes, which are comfortably larger than the observed inter-layer separations in the regions within \(7'\) of the Trapezium that are included in Figure~\ref{fig:spectrophotometry} of this paper. 

Pre-attenuation is the reduction of the flux \(F_0\) incident on the scattering layer due to absorptions in material at smaller radii that does not contribute to the observed scattered intensity \(I\). 
In the brightest regions of the  nebula it can be shown that the majority of the emission in ionized lines such as \hb{} arises in a relatively thin layer near the ionization front (e.g., \citealp{1995ApJ...438..784W}), but there is also a more extended diffuse component to the emission, which becomes relatively more important at greater distances \citep{Baldwin.J91Physical-conditions-in-the-Orion-Nebula, 2005ApJ...627..813H}. 
This pre-attenuation will affect the line ratios and equivalent widths only if it is \textit{differential}, that is, affecting one scattering process more than another. 
For incident radiation in the optical and far-ultraviolet bands, the dominant absorption process is always due to dust, whereas for ionizing extreme-ultraviolet radiation it may be dust or hydrogen, depending on the local ionization parameter.  
In \S~\ref{sec:relat-import-dust} below it is shown that, in the diffuse ionized gas responsible for the pre-attenuation, dust is the dominant opacity source in the EUV band also.
Since the dust absorption cross section is very similar at optical and EUV wavelengths (Fig.~19 of \citealp{Baldwin.J91Physical-conditions-in-the-Orion-Nebula}), pre-attenuation will have almost no effect on \EWhb, which is sensitive to the EUV/optical flux ratio. 
The dust absorption cross section in the FUV band is about 20--50\% higher than in the visual band, so that pre-attenuation may affect \EWni, which is sensitive to the FUV/optical flux ratio. 
However, the total continuum optical depth to the [\ni]-scattering layer is only of order unity (see Fig.~\ref{fig:EmissivityVsDepth}) and the optical depth of any diffuse pre-attenuating gas must be substantially less than this, so the effect is likely to be small.

Limb-brightening is the increase in intensity of the emergent intensity as the viewing angle becomes more closely edge-on, due to the increased optical path through the scattering layer.  
In a strict plane-parallel approximation, the limb brightening does not saturate until the scattering layer is optically thick to the emergent radiation along the viewing direction, but any curvature in the layer will impose an additional limit on the degree of limb-brightening. 
This arises since the maximum path length through the layer is approximately \(2\sqrt{2R h}\), where \(R\) is the radius of curvature and \(h\) is the layer thickness, meaning that the maximum boost that limb-brightening can give the emergent intensity with respect to the face-on (\(\mu = 0\)) value is of order \(2\sqrt{2R/h}\).  
Typical values of \(R/h\) vary from \(\simeq 3.5\) for the \hb-scattering layer to \(\simeq 100\) for the [\ni]-scattering layer, giving maximum boost factors of \(\simeq 5\) and \(\simeq 30\), respectively. 
Since optical depth effects also limit the boost factor to a maximum of about 5 (see \S\S~\ref{sec:albedo-hb} to \ref{sec:albedo-ni} above), the extra limiting of limb-brightening by curvature effects will be unimportant, except arguably for \hb{}.

\subsubsection{Relative importance of dust versus hydrogen opacity at EUV wavelengths }
\label{sec:relat-import-dust}
By using the equation of local photoionization equilibrium to rewrite the hydrogen photoabsorption rate in terms of the recombination rate, it is straightforward to show that dust will dominate the EUV opacity in ionized gas for densities less than \(n' = F_0 \sigma\dust / \alpha\B\), where \(\sigma\dust\) is the EUV dust absorption cross-section. 
Taking \(\sigma\dust = 5 \times 10^{-22} \U{cm^{-2}\ H^{-1}}\) (Fig.~19 of \citealp{Baldwin.J91Physical-conditions-in-the-Orion-Nebula}) and using the ionizing luminosity of the Trapezium stars listed in Table~\ref{tab:stars}, one finds \(n' \simeq (7000 / D^{-2})~\pcc\), where \(D\) is the projected distance from the Trapezium in arcminutes (assumed to be on average \(\sqrt{3}/2\) times smaller than the true distance). 
Coincidentally, this equation for \(n'\) is very close to the reference line drawn on Fig.~6 of \citet{ODell:2010}, which shows observationally derived electron densities as a function of distance, and from which it can be seen that \(n < n'\) for \(D < 2'\) but that \(n \sim n'\) for \(D = 2'\)--\(7'\).  
The diffuse ionized gas in the interior of the \hii{} region is likely to have somewhat lower density than the mean densities derived from line ratios. 
Therefore, we conclude that dust is the dominant opacity source for EUV radiation in the diffuse ionized gas at all radii covered by our observations. 

Note, however, that this does not mean that dust is the dominant EUV opacity source in the \hii{} region as a whole.  In fact, only 10--20\% of the ionizing photons are absorbed by dust, but the hydrogen absorption is weighted towards the edge of the \hii{} region, rather than the diffuse interior gas.

\section{Non-thermal line broadening of [\ni] and other lines}
\label{sec:non-thermal-line}
\label{sec:boldm-broad-ni}

\begin{table*}[tp]
  \centering
  \caption{Non-thermal line widths of different gas phases in Orion}
  \label{tab:linewidths}\small
\begin{tabular}{l@{}cccccccc}\hline
 & Atomic & Mean    & Gas         & Sound & \multicolumn{3}{@{}c}{\dotfill Line width FWHM\dotfill } & Mach \\ 
 & Weight & mass    & temperature & speed & Total & Thermal & Non-thermal       & number \\
Species& \(A\) & \(\mu\) & \(T\) (\K) & \( c_{\mathrm{s}}\)  (\kmps)& \(\delta V\) (\kmps)& \(\delta V_{\mathrm{th}}\) (\kmps)& \(\delta V_{\mathrm{nth}}\) (\kmps)& \(M\) \\
\hline
CO & 28 & 2.36 & \(\phn\phn 40 \pm 10\phn   \) & \(\phn 0.37 \pm 0.05 \) & \(\phn 3.0 \pm 0.5   \) & \(0.26 \pm 0.03 \) & \(\phn 2.99 \pm 0.50 \) & \(4.0 \pm 0.9 \) \\
{}[\ion{C}{2}] & 12 & 1.30 & \(\phn 400 \pm 100 \) & \(\phn{}1.59 \pm 0.20 \) & \(\phn{}4.0 \pm 1.0 \) & \(1.23 \pm 0.15 \) & \(\phn 3.81 \pm 1.05 \) & \(1.2 \pm 0.4 \) \\
{}[\ion{N}{1}] & 14 & 1.30 & \(2000 \pm 500\) & \(\phn{}3.56 \pm 0.45 \) & \(\phn{}6.0 \pm 3.0 \) & \(2.56 \pm 0.32 \) & \(\phn 5.43 \pm 3.32 \) & \(0.8 \pm 0.5 \) \\
{}[\ion{O}{1}] & 16 & 0.90 & \(9500 \pm 500\) & \(\phn{}9.33 \pm 0.25 \) & \(12.6 \pm 2.4\) & \(5.21 \pm 0.14 \) & \(11.47 \pm 2.64\) & \(0.6 \pm 0.1 \) \\
{}[\ion{N}{2}] & 14 & 0.68 & \(9000 \pm 500\) & \(10.45 \pm 0.29\) & \(16.3 \pm 3.5\) & \(5.42 \pm 0.15 \) & \(15.37 \pm 3.71\) & \(0.7 \pm 0.2 \) \\
{}[\ion{O}{3}] & 16 & 0.65 & \(8400 \pm 500\) & \(10.33 \pm 0.31\) & \(15.5 \pm 4.8\) & \(4.90 \pm 0.15 \) & \(14.71 \pm 5.06\) & \(0.7 \pm 0.2 \) \\
\hline
\end{tabular}

\tablerefs{
  CO: \citet{Wilson:2001}; [\ion{C}{2}]: \citet{Boreiko:1988}; other lines: \citet{Baldwin.J00High-Resolution-Spectroscopy-of-Faint-Emission, 2008RMxAA..44..181G}
}
\tablecomments{
  All linewidths have been corrected for instrumental broadening.
}

\end{table*}

In the Orion Nebula, as elsewhere in the interstellar medium, significant non-thermal linewidths are observed to be ubiquitous in all gaseous phases \citep{2001ARA&A..39...99O, 2003AJ....125.2590O}.  
This is shown in Table~\ref{tab:linewidths}, which collates measurements from the literature of linewidths \(\delta V\) and gas temperature \(T\) for various emission lines in the central Orion Nebula, ranging from fully molecular to fully ionized species.  
The expected thermal FWHM is \(\delta V_{\mathrm{th}} = 0.214 \sqrt{(T / A)}~\kmps\) where \(T\) is the temperature in \K{} and \(A\) is the atomic weight of the emitting species in units of the proton mass \(m_{\mathrm{p}}\).  
This is subtracted in quadrature from the total width to give the non-thermal broadening component: \(\delta V_{\mathrm{nth}} = (\delta V^2 - \delta V_{\mathrm{th}}^2)^{1/2} \).  
It can be seen that the non-thermal component dominates over the thermal in all cases and increases in magnitude from about 3~\kmps{} in fully molecular gas up to about 15~\kmps{} in the fully ionized gas.  
If the non-thermal broadening is truly due to gas motions, then an approximate characteristic Mach number of these motions can be calculated as \(M = 0.5\, \delta V_{\mathrm{nth}} / c_{\mathrm{s}}\), where \(c_{\mathrm{s}} = \sqrt{(k T / \mu m_{p})}\) is the sound speed and \(\mu\) is the mean mass per particle.  
This Mach number is shown in the last column of the Table, and in contrast to the linewidth it \emph{decreases} with increasing ionization of the gas: the non-thermal motions are highly supersonic in fully molecular gas, slightly supersonic in the PDR, and slightly subsonic in the ionized gas. 

Optical and infrared emission lines from \hii{} regions and PDRs, such as the majority of those listed in Table~\ref{tab:linewidths} are usually optically thin.  
Therefore, the observed line widths give no information about the spatial scales at which the broadening mechanism operates.  
On the other hand, for optically thick lines, such as the FUV lines that are responsible for pumping the [\ni{}] emission, one can divide potential broadening mechanisms into two categories: \emph{microscopic} and \emph{macroscopic}, according to whether they occur at scales that are smaller than or larger than the relevant photon mean free path.  
Of the two, only microscopic mechanisms act to broaden the absorption profile and so affect the radiative transfer of the line, whereas macroscopic mechanisms simply act to broaden the emergent intensity profile.  

The line broadening that we derive for the pumping lines in order to explain the observed optical [\ni{}] line brightness (see Figure~\ref{fig:varyTurb}) is 
similar to, but smaller than, the broadening observed in the lines themselves (Table~\ref{tab:linewidths}), implying that the non-thermal broadening mechanism must be microscopic in nature. 
In other words, it should occur on scales of less than \(10^{14}~\mathrm{cm}\), which is the approximate mean free path of the \(954~\AA\) pumping line.  

In the \hii{} region, transonic turbulence is expected to be driven at the scales of photoevaporation flows from dense globules and filaments \citep{2006ApJ...647..397M, Arthur:2011, Ercolano:2011a}.  
The most vigorous photoevaporation flows only occur at scales larger than about 10\% of the \hii{} region radius \citep{2003RMxAC..15..175H}.  
In the Orion nebula, the closest approach of the ionization front to the ionizing stars is about \(0.2\)~parsec \citep{1995ApJ...438..784W}, so we assume that turbulent velocities of amplitude \(11~\kmps\) are present at scales of \(0.02\)~parsec, or \(6\times 10^{16}~\mathrm{cm}\).  
In a  Kolmogorov-type turbulent energy cascade, velocity differences scale with separation \(\ell\) as \(\delta v \sim \ell^{1/3}\).  
Therefore, on the scale of the \ni{} mean free path the turbulent broadening should be only approximately \(1~\kmps\), which is much smaller than our derived microscopic broadening, which means that this is not the mechanism we seek. 

A further potential source of broadening is the systematic acceleration of the gas as it is dissociated, heated, and ionized by the advancing front.  
However, the [\ni] emission arises in regions where the gas is still predominantly neutral, with temperature ranging from 1000 to 5000~\K{} (see Fig.~\ref{fig:EmissivityVsDepth}), for which the velocity increase is expected to be small since the greater part of the gas acceleration occurs in the warmer, partially ionized zone where the [\oi] lines arise.  
For example, a plane-parallel model of a D-critical ionization front (Eqs.~(A5)--(A8) of \citealp{2005ApJ...621..328H}) implies a total broadening FWHM for the [\ni] lines of less than \(2~\kmps\) by this process. 

We therefore see that none of the broadening mechanisms that have been succesfully invoked to explain the observed widths of neutral and ionized collisional lines are successful in explaining the observed characteristics of the fluorescent [\ni] lines.  
Other potential mechanisms such as instabilities of the ionization front itself \citep{2002MNRAS.331..693W, 2008ApJ...672..287W} are not promising either, since they are unlikely to produce microturbulence at a sufficiently small scale.  
Neither can broadening due to dust scattering in the neutral veil \citep{1998ApJ...503..760H} be the explanation, since this would have no affect on the radiative transfer of the \ni{} pumping lines.

A more promising mechanism for generating turbulent velocities on a very small scale is the action of thermal instabilities \citep{Koyama:2002} in the shocked neutral layer that precedes the ionization front.  
It is suggestive that the temperature range over which the [\ni] lines form in the PDR (1000--3000~\K) is similar to the range over which the ISM is known to be thermally unstable \citep{Field:1969}.
If such an instability were to occur in the PDR, then the smallest scale at which fragments could arise (and hence turbulence be driven) is given by the Field length \(\lambda_\mathrm{F}\) \citep{Field:1965}, which represents the scale below which temperature fluctuations will be smoothed out by thermal conduction.  
Assuming saturated conduction by free electrons \citep{ZelDovich:1967}, one finds a value of \(\lambda_\mathrm{F} \simeq (10^{15} / n) \mathrm{\ cm} \), which is roughly 100 times smaller than the thickness of the [\ni] pumping layer. 
The role of thermal instability in generating the required microscopic non-thermal broadening therefore merits further investigation.

\end{appendix}

\end{document}